\documentclass[twocolumn,prb]{revtex4}

\usepackage{graphicx}
\usepackage{dcolumn}
\usepackage{bm}
\usepackage{epsfig}

\newcommand{\e}[1]{\epsilon_{\vec{k}}}

\newcommand{\ba}{\begin{eqnarray}}
\newcommand{\ea}{\end{eqnarray}}

\newcommand{\lv}{Liouvillian\ }
\newcommand{\lvs}{Liouvillians\ }
\newcommand{\mar}{\marginpar}
\newcommand{\nl}{$N_L$}

\def\8{\infty}

\def\undertext#1{\vtop{\hbox{#1}\kern 1pt \hrule}}

\def\be{\begin{equation}}
\def\ee{\end{equation}}
\def\bea{\begin{eqnarray} & &}
\def\eea{\end{eqnarray}}

\begin{document}
\title{A Farewell to Liouvillians}

\author{Vadim Oganesyan$^{1}$}
\author{J. T. Chalker$^{2}$}
\author{S. L. Sondhi$^{1}$}

\affiliation{\mbox{$^{1}$}Deptartment of Physics,
Princeton University,
Princeton, NJ 08544, USA}
\affiliation{\mbox{$^{2}$}
Theoretical Physics, University of Oxford, 1 Keble Road, Oxford,
OX1 3NP, United Kingdom}

\date{\today}

\begin{abstract}

We examine the Liouvillian approach to the
quantum Hall plateau transition, as introduced recently by
Sinova, Meden, and Girvin [Phys. Rev. B {\bf 62}, 2008 (2000)]
and developed by Moore, Sinova and Zee [Phys. Rev. Lett. {\bf 87},
046801 (2001)]. We show that, despite appearances to the contrary,
the Liouvillian approach is not specific to the quantum mechanics
of particles moving in a single Landau level: we formulate it for
a general disordered single-particle Hamiltonian.  We next examine
the relationship between Liouvillian perturbation theory and
conventional calculations of disorder-averaged products of Green
functions and show that each term in Liouvillian perturbation theory
corresponds to a specific contribution to the two-particle Green
function. As a consequence, any Liouvillian
approximation scheme may be re-expressed in the language of Green
functions. We illustrate these ideas by applying Liouvillian
methods, including their extension to $N_L > 1$ Liouvillian flavors,
to random matrix ensembles, using numerical calculations for small integer
$N_L$ and an analytic analysis for large $N_L$.
We find that behavior at $N_L > 1$ is different in 
qualitative ways from that at $N_L=1$.
In particular, the $N_L = \infty$ limit
expressed using Green functions
generates a pathological approximation, in which two-particle
correlation functions fail to factorize correctly at large
separations of their energy,
and exhibit spurious singularities inside the band of random matrix energy
levels. We also consider the large $N_L$
treatment of the quantum Hall plateau transition,
showing that the same undesirable features are
present there, too.

\end{abstract}
\maketitle


\section{Introduction}

It has long been appreciated that the quantum Hall effect depends
for its existence on localization by disorder of quasiparticles  
in Landau level tails, and that neighboring quantum Hall plateaus 
are separated by a continuous quantum phase transition characterized by
a diverging localization length.\cite{Huckestein} 
The scaling ideas which encapsulate this understanding
\cite{Pruisken,Khemlnitskii} are supported by 
extensive numerical studies \cite{Huckestein} and by some
of the available experimental data\cite{Wei,Haug} although the
full experimental situation remains unsettled \cite{shahar}.
An analytic theory of the transition, however, has proved elusive, 
even for the
simplest models which include disorder and magnetic field but omit
electron-electron interactions.

In this context, two successive recent developments have attracted 
interest: Sinova, Meden and Girvin\cite{SMG} (SMG)
have introduced a Liouvillian approach to localization in the 
lowest Landau level, which Moore, Sinova and Zee\cite{MZS} (MSZ)
have extended, with the introduction of $N_L$ Liouvillian flavors
and an expansion in powers of $1/N_L$ about 
the large $N_L$ limit\cite{GurZee}. In brief, the Liouvillian is the time evolution
operator for electron probability density, and the information it encodes
on localization is integrated over states at all energies within
the disorder-broadened Landau level. Despite the energy integration,
the critical behavior of the localization length at the plateau 
transition can, in principle, be extracted from the dependence of the 
Liouvillian propagator on frequency $\omega$, provided this is known 
sufficiently accurately 
(see SMG 
and also Appendix A of
this paper).  The original work of SMG reported a calculation of the Liouvillian
 propagator using a version of the self-consistent Born approximation
 (SCBA), which is exact for the  
$N_L=\infty$ limit of MSZ, and yields diffusive time evolution without 
localization. MSZ have conjectured that the ${\cal O}(1/N_L)$ correction, 
which 
is logarithmically divergent at small $\omega$, 
may be used to arrive at an estimate for the localization 
length exponent $\nu$.

{\it Prima facie} there are three reasons to think that this work
may have outflanked the obstructions faced by previous approaches.
First, the derivation of the \lv formalism by SMG invokes the
algebra of density operators projected onto the lowest Landau
level, thus appearing to build in the physics of high magnetic
fields at the first step. Second, the formalism deals directly
with a disorder-averaged two-particle quantity, avoiding
intermediate calculations of a one-particle Green function,
and hence, plausibly, goes directly to the heart of the 
matter. Third, the extension to $N_L$ \lv flavors is distinct
from the standard extension to $N$ orbitals in localization
problems, \cite{oppweg} which is known to capture weak localization
physics but not the quantum Hall delocalization transition\cite{fn-Ntransitions}.
This gives grounds for hope that the logarithm found by MSZ
may not simply be due to weak localization effects and might indeed
be used to get an estimate for the correlation length exponent.

Given the potential importance of these developments, it seems
useful to investigate the Liouvillian approach in some detail
and to relate it to established methods. This is our objective in the
present paper.
There are four distinct facets to our results. 
First, we show that although the Liouvillian approach has appeared in
past work to be tailored specifically to the quantum mechanics of
particles moving in a single Landau level, since it makes use of
the algebra of projected density operators, it can in fact be formulated
for any disordered single-particle Hamiltonian. Second, we
compare the perturbation expansion for the Liouvillian
propagator with conventional calculations of the disorder-averaged
two-particle Green function. We demonstrate
that each term in the perturbation expansion for the Liouvillian
propagator corresponds to a specific combination of terms
in the Green function expansion. As a consequence, any approximation
scheme within one approach has an equivalent in the other approach.
Moverover, it is possible to translate between the two approaches
in {\it either} direction: to go from the two-particle Green 
function to the Liouvillian simply involves an energy integration, while a
more elaborate proceedure, which we set out, is required to undo this 
energy integration
and pass in the opposite direction. Third, we illustrate
these ideas by applying the Liouvillian approach to random matrix
theory, discussing the Gaussian unitary and orthogonal ensembles.
We obtain an analytic solution in the limit of large flavor number $N_L$,
for arbitrary matrix size $N$, and we supplement this with numerical
calculations for finite $N_L$, finding qualitative differences between
results at $N_L = \infty$ (and all $N_L>1$) 
and those at $N_L=1$. We also undo the energy
average at $N_L=\infty$, showing that the Liouvillian SCBA has a very
different character from established approximations when translated
into a calculation for the two-particle Green function. It constitutes
an approximation without the usual structure based on single-particle 
self-energies and two-particle irreducible vertices. Disappointingly,
this is not progress as the approximation is pathological: the
resulting Green functions exhibit spurious singularities 
{\it inside the band}
and fail to factorize correctly at large energy separations. Finally,
we return to the plateau transition. Progress in this case is more 
difficult, because even the Liouvillian SCBA of SMG requires a numerical 
solution. We are nevertheless able to show that the two-particle Green 
function generated by the Liouvillian SCBA has undesirable features
in this case, too,
failing to factorise as it should for large separations between 
pairs of its spatial arguments, and exhibiting spurious singularities
as a function of energy in this limit.

In sum, on one hand we have shown generally that any Liouvillian 
approximation can equally well
be expressed using the better-understood machinery of Green functions,
and on the other hand we have argued that the only existing
basis for Liouvillian calculations, the $1/N_L$ expansion, 
is seriously flawed. 
In combination, these results leave us pessimistic about the scope for
advances in the theory of the quantum Hall plateau transition
using Liouvillian methods. 

In the balance of the paper we set out the technical content of
these assertions. In Section II we develop \lv machinery in a
general setting and describe, first, how to go from two particle
Green functions to \lvs by energy integration, and second, our algorithm
for undoing this energy integration within a given \lv approximation. In
Section III we review the $1/N_L$ expansion scheme introduced
by MSZ, again in a general setting. Section IV is devoted to
a detailed examination of the \lv technique applied to the
test case of random matrix statistics. In Section V we consider
the quantum Hall problem and describe the pathology that
is immediately apparent by recourse to the previous results.
We end with a summary and 
three appendices which provide: 
a more careful discussion of the critical behavior of the
quantum Hall \lv than is available in previous work; some
details omitted in the main text; and a construction that
yields an algebra for an arbitrary single-particle
Hamiltonian that is identical to the algebra of
density operators projected onto the lowest 
Landau level. 

\section{General considerations} 
\label{sec:gendef}
\subsection{From two-particle Green functions to the Liouvillian}

Consider a single-particle Hamiltonian $\hat{H}$ acting on 
basis states $|a\rangle$ in a space $\cal V$
of dimensionality $N$. Although all of its properties are encapsulated in
the corresponding  one-particle Green functions
\ba
G^{\pm}_{ab}(E)&=&\langle a|\frac{1}{E-\hat{H}\pm i\delta} |b \rangle,
\label{eq:G}
\ea
(where $\delta$ is a positive
infinitesimal), in most problems of interest their more easily computed
disorder averages $\langle G^{\pm} \rangle$ have little of the interesting 
information 
contained in $G^{\pm}$.
One is then forced to consider higher order correlators such as the
two-particle retarded-advanced Green function 
\be
K^{+-}_{aA;bB}(E, \omega)=
G^{+}_{ab}(E+\frac{\omega}{2})G^{-}_{BA}(E-\frac{\omega}{2}),
\label{eq:K}
\ee
whose disorder average does contain useful information.
Analogously one can define $K^{++}$,$K^{-+}$ and $K^{--}$.
Here and in the following, we use $\pm$ as superscripts 
to indicate retarded and advanced Green functions, and reserve lower
case state labels ($a,b$) for the former, and upper case ones ($A,B$)
for the latter.

The central object in this paper is the energy integral 
of $K^{+-}_{aA;bB}(E,\omega)$
\be
\Pi_{aA;bB}
(\omega)\equiv
\int_{-\infty}^{\infty}\frac{d E}{2\pi i}\  
{K}^{+-}_{aA;bB}
(E,\omega).
\label{pidef}
\ee

We will now show that, quite generally, this can be expressed as
the {\it one particle}
Green function of a Liouville ``super-operator''
(the Liouvillian) which itself acts on the space $\cal A_{\cal V}$ 
of all linear
operators on the space $\cal V$. Clearly, this Liouvillian Green function
is to be distinguished from the usual Green function of the Hamiltonian,
introduced in Eq.\,(\ref{eq:G}).
The space $\cal A_{\cal V}$ is $N^2$ dimensional and is spanned by the 
basis set $ |a\rangle \langle b|$.
To emphasize that these operators themselves form a 
linear space, we will use the notation $|a,b) = |a\rangle \langle b|$.
A natural inner product on this space is defined as ${\rm Tr} P^\dagger
Q$ for any operators $P$ and $Q$ that belong to $\cal A_{\cal V}$. 
In terms of the basis this yields
\begin{equation}
(a, b|c,d) \equiv {\rm Tr}\{ [|a\rangle \langle b|]^{\dagger}
|c\rangle \langle d|\} =
\langle a | c \rangle \langle d | b\rangle\,.
\end{equation}
Finally, pairs of operators $P$, $Q$ define a super-operator 
on $\cal A_{\cal V}$ via 
the action $P \circ Q |a,b) = P |a\rangle \langle b| Q$ from left and right
respectively.

Now consider rewriting Eq.\,(\ref{pidef}) (with tildes denoting
 inverse Fourier transforms)
\begin{widetext}
\ba
\Pi_{aA;bB}(\omega)&=&
\int_{-\infty}^{\infty}\frac{d E}{2\pi i}\  
G^+_{ab}(E+\frac{\omega}{2}) \,
G^-_{BA}(E-\frac{\omega}{2}) \nonumber \\
&=&\int_{-\infty}^{\infty}
\frac{d E}{2\pi i} 
\int_{-\infty}^{\infty} dt_1 
\int_{-\infty}^{\infty} dt_2
 e^{i(E + \omega/2+i \delta)t_1+i(E -\omega/2-i \delta)t_2}
\tilde{G}^{+}_{ab}(t_1) \,
\tilde{G}^{-}_{BA}(t_2) \nonumber\\
&=&
-i \int_{-\infty}^{\infty} dt \, e^{i (\omega +i \delta) t} \tilde{G}^{+}_{ab}(t) \,
\tilde{G}^{-}_{BA}(-t)  
=
-i \int_{-\infty}^{\infty} dt e^{i (\omega +i \delta) t} (\theta(t))^2
\langle 
a|e^{-iHt}| b \rangle \langle B|e^{iHt} |A \rangle \nonumber 
\\
&=& 
-i \int_{-\infty}^{\infty} dt e^{i (\omega +i \delta) t} \theta(t) 
Tr\{(e^{iHt}|A\rangle\langle a|e^{-iHt})|b\rangle\langle B| \}
\equiv
-i \int_{0}^{\infty}dt e^{i (\omega +i \delta) t}\ {_t}( a,A| b,B)_0
\ea 
\end{widetext}
The single particle Hamiltonian induces a time evolution super-operator,
the \lv $\cal L$, on $\cal A_{\cal V}$
\be
|a,A)_t = e^{i {\cal L} t} |a,A)_0
\ee
and the matrix elements of $\cal L$ can be read off from
\begin{widetext}\be
\frac{d}{dt} |a\rangle \langle A|= i [\hat{H},|a\rangle \langle A|]
= i \sum_{bB}
( H_{ab} \delta_{BA} - \delta_{ab} H_{BA}) |b\rangle \langle B|
\equiv i \sum_{bB}\ {\cal L}_{aA;bB} |b\rangle \langle B| \ .
\label{L}
\ee
\end{widetext}
More abstractly,
\be
\hat{\cal L}=\hat{H} \circ \hat{1}- \hat{1} \circ \hat{H} \ .
\label{eq:L}
\ee
In terms of $\cal L$ we have finally 
\ba
\Pi_{aA;bB}(\omega)&=& (a, A |\frac{1}{\omega-\hat{\cal L} + i \delta}
| b,B) 
\label{eq:Pi}
\ea
{\rm
which is the promised rewriting of $\Pi_{aA;bB}(\omega)$. While 
Eq.\,(\ref{eq:L}) does not appear in the work of SMG and MSZ, 
our $\cal L$ is simply the generalisation for arbitrary $\hat H$
of the Liouvillian discussed for the quantum Hall plateau 
transition in previous papers.
We continue our general discussion below and take up the 
plateau transition in Section \ref{sec:QH}.

\subsection{Exact properties of $\hat{\cal L}$}

A few general and exact properties of $\hat{\cal {L}}$ can be inferred
from the preceding formulae.
The eigenvalues $\lambda_{ab}$ of $\hat{\cal {L}}$ are simply related 
to the eigenvalues $\epsilon_a$ of $\hat{ H}$ by
\be
\lambda_{ab}=\epsilon_a-\epsilon_b.
\label{eq:lee}
\ee
This implies that $\lambda_{ab}=-\lambda_{ba}$, so that the
eigenvalues of $\hat{\cal L}$ occur in pairs symmetric about $\lambda=0$. 
In addition, the \lv has at least $N$ zero eigenvalues, since 
$\lambda_{mm}=0$. 
Finally, if the eigenvalues of ${\hat H}$ occupy a band of width $W$, 
the bandwidth 
of ${\hat{\cal L}}$ is $2W$.

\subsection{Perturbation theory}

From Eq.\,(\ref{eq:Pi}) we can generate the perturbation expansion
\begin{equation}
{\Pi}_{aA;bB}
(\omega)=
\sum_{n=0}^\infty 
 \frac{ (a,A| \hat{{\cal L}}^n |b,B)}{ (\omega + i \delta)^{n+1}} \, .
\label{eq:Pi(n)}
\end{equation}
It is interesting to see how this arises from energy 
integrating the perturbative
expression for the two-particle Green function.
Consider the contribution to 
$K^{+-}_{aA;bB} (\omega)$ at order $n$ in $\hat{H}$, which is
\be
K^{(n)}_{aA;bB}(E,\omega)=
\sum_{m=0}^n
\frac{[\hat{H}^m]_{ab} [\hat{H}^{n-m}]_{BA}}{(E-\frac{\omega}{2}-i\delta)^{m+1}
(E+\frac{\omega}{2}+i\delta)^{n+1-m}}\,\,.
\label{eq:K(n)}
\ee
Integration over $E$,
\begin{widetext}
\begin{equation}
\int_{-\infty}^{\infty}K^{(n)}_{aA;bB}(E,\omega)\frac{dE}{2\pi i} = 
\frac{1}{(\omega + i \delta)^{n+1}}
\sum_{m=0}^n \frac{(-1)^m n!}{(n-m)!m!}
[\hat{H}^m]_{ab}[\hat{H}^{n-m}]_{BA} \,\,,
\label{eq:intK(n)}
\end{equation}
\end{widetext}
produces precisely the corresponding term in the 
expansion for $\hat{\Pi}(\omega)$.

\subsection{Disorder averaging}
For the Green function perturbation theory
with Gaussian randomness in $\hat H$,
the effect of disorder averaging is to replace
$\langle \hat{H}^{2n} \rangle$ with a sum of products of all pairwise
contractions $\langle \hat{H} \hat{H} \rangle$. 
We can translate this into the language of 
the \lv theory by using the definition of
$\hat{\cal{L}}$ to express its matrix elements in terms of those of
$\hat{H}$,
then using the correlators for the matrix elements of $\hat{H}$ 
to obtain those for $\hat{\cal{L}}$, 
\begin{widetext}
\ba
\langle {\cal L}_{aA;bB}\ {\cal L}_{cC;dD}\rangle &=&
\langle (\delta_{ab}
H_{BA}-H_{ab}\delta_{BA})(\delta_{cd}H_{DC}-H_{cd}\delta_{DC})\rangle 
\nonumber \\
&=&
\delta_{ab} \delta_{cd} \langle H_{BA} H_{DC} \rangle
-\delta_{ab} \delta_{DC} \langle H_{BA} H_{cd} \rangle
-\delta_{BA} \delta_{cd} \langle H_{ab} H_{DC} \rangle
+\delta_{BA} \delta_{DC} \langle H_{ab} H_{cd} \rangle\,, 
\label{LH}
\ea
\end{widetext}
and finally using these to average over the powers $\hat{\cal L}^{2n}$
that occur in the perturbative expansion.

\subsection{From \lv perturbation theory to Green function perturbation theory}
\label{subsec:unint}
The purpose of this subsection is to construct an algorithm for
passing from a given (likely approximate) expression 
for ${\hat{\Pi}}(\omega)$ to one for
${\hat K}(E,\omega)$. 
Specifically, we would like to associate uniquely each disorder
averaged diagram for ${\hat{\Pi}}(\omega)$ with a corresponding set involving 
Green functions.
To make the reader's task simpler we we first summarize the basic idea
of the algorithm,
then provide a representative example, and finally state
the general formula. 

\subsubsection{Basic idea}

In going between Eq.\,(\ref{eq:K(n)}) and Eq.\,(\ref{eq:intK(n)}), we
have lost 
the distinction between advanced and retarded propogators (since $E\pm
\frac{\omega}{2} \pm i\delta$ 
have been traded for $\omega$) but the matrix elements still record
where the original disorder lines were attached in the two-particle
Green function diagram. If we reconstruct
this information from the \lv matrix elements, where it must reside
since they are defined by matrix elements of the Hamiltonian, our
remaning task is merely one of undoing the signs and combinatorial
factors introduced by the energy integration. This is easier done
than said.\cite{Iddo}

\subsubsection{Example}
Let us see how this works at lowest non-trivial order (fourth, since at second
order there is only one \lv diagram). Of the three diagrams in
Fig.\,\ref{fig:fourthorder}
\begin{figure}[h]
\epsfxsize=2.5truein
\centerline{\epsffile{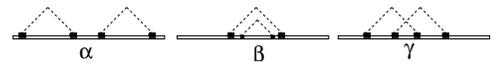}}
\caption{
Diagrams for $\hat{\Pi}(\omega)$ at fourth order in $\cal L$.}
\label{fig:fourthorder}
\end{figure}
we will use the middle one, denoting its contribution
to $\Pi_{aA;bB}(\omega)$ by  $\Pi_{aA;bB}^{(\beta)}(\omega)$.
In terms of $\hat H$ it is (summation over all repeated indices is
implied)
\begin{widetext}
\ba
\label{eq:Pi(4beta)}
\Pi^{(\beta)}_{aA;bB}(\omega)&=&
\frac{1}{(\omega+i\delta)^5}\langle {\cal L}_{aA;uU}{\cal
L}_{wW;bB}\rangle \langle  
{\cal L}_{uU,vV}{\cal L}_{vV;wW}\rangle 
\label{eq:elelelel}
\\
&=&\frac{1}{(\omega+i\delta)^5}[\delta_{ab}\langle H_{BW}H_{UA}
\rangle\langle \hat{H}^2\rangle_{WU}+
\delta_{BA}\langle H_{au}H_{wb}\rangle\langle \hat{H}^2\rangle_{uw}
\nonumber \\
&&-2
\langle H_{au}H_{vb}\rangle \langle H_{BA}H_{uv}\rangle
-2
\langle H_{ab}H_{VU}\rangle \langle H_{BV}H_{UA}\rangle
\nonumber\\
&&-
\langle H_{au}H_{BA}\rangle \langle H^2\rangle_{ub}
-
\langle H_{ab}H_{BU}\rangle \langle H^2\rangle_{UA}
-
\langle H^2\rangle_{au} \langle H_{ub}H_{BA}\rangle
-\langle H_{ab}H_{UA}\rangle \langle H^2\rangle_{BU}
\nonumber \\
&&
+2 \langle \hat{H}^2\rangle_{ab}\langle \hat{H}^2\rangle_{BA}+4\langle H_{au}H_{BU}\rangle \langle H_{ub}H_{UA}\rangle]\,\,,
\label{eq:eltohhhh}
\ea
\end{widetext}
which can be represented graphically by drawing the bare \lv
propagator as a double line and resolving disorder vertices using the
Hamiltonian as indicated in Fig.\,\ref{fig:LLLLtoHHHH}.
This is done in detail in Fig. \ref{fig:HHHH}, in which the first
diagram
is Eq. (\ref{eq:elelelel}) and last ten are 
Eq. (\ref{eq:eltohhhh}).
\begin{figure}[h]
\epsfxsize=2.5truein
\centerline{\epsffile{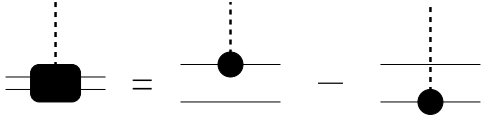}}
\caption{
Graphical representation of 
$\hat{\cal L}=\hat{H} \circ \hat{1}-\hat{1} \circ\hat{H}$.
}
\label{fig:LLLLtoHHHH}
\end{figure}
\begin{figure}[h]
\epsfysize=5truein
\centerline{\epsffile{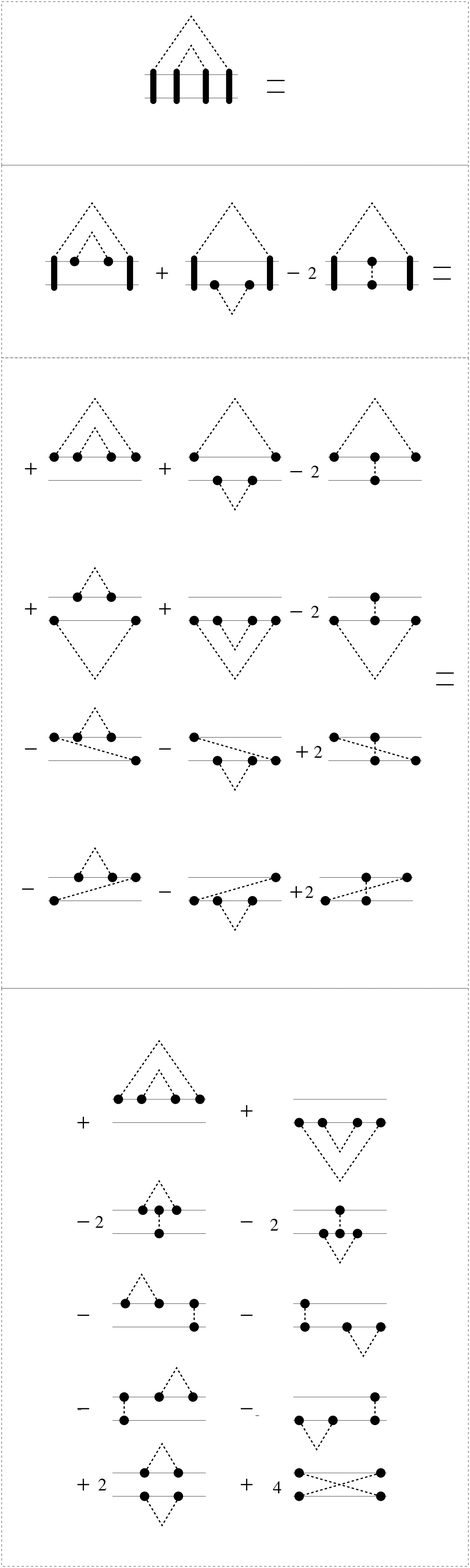}}
\caption{Graphical representation of the correspondence 
between \lv and Green function perturbation series.
First diagram corresponds to Eq. \ref{eq:elelelel} and contains two
\lv contractions. Next, three diagrams are obtained by redrawing one of
the contractions using Fig. \ref{fig:LLLLtoHHHH}, each of which is in turn
decomposed by repeating the procedure on the remaining \lv
contraction.
Finally, the last 
ten diagrams represent Eq. \ref{eq:eltohhhh}.
}
\label{fig:HHHH}
\end{figure}
\begin{figure}[h]
\epsfysize=3truein
\centerline{\epsffile{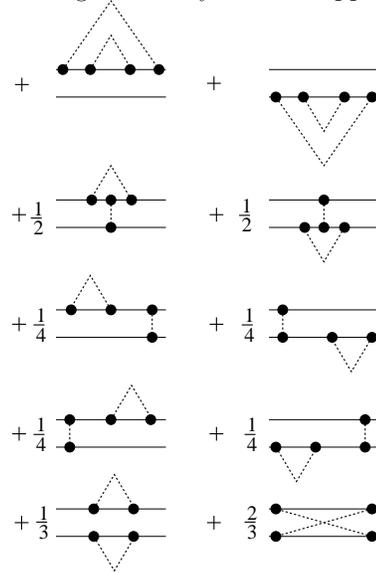}}
\caption{
Green functions for the Liouvillian diagram $\beta$ of
Fig. \ref{fig:fourthorder}, represented by
Eqns.\,(\ref{eq:eltohhhh}) and (\ref{eq:K(4beta)}). This figure should be compared with the last ten diagrams of  Fig. \ref{fig:HHHH}. }
\label{fig:Hresolved}
\end{figure}
The final step is to undo the signs and combinatorial factors generated by the
energy integration in going from Eq.\,(\ref{eq:K(n)}) to
Eq.\,(\ref{eq:intK(n)}). For each term in our example, this may be done 
by simply counting how many times $H$ appears on
upper or lower lines, representing retarded or advanced Green functions.
In this way, 
writing $E_\pm \equiv E\pm{\omega}/{2}\pm i\delta$,
we find that Eq.\,(\ref{eq:Pi(4beta)}) generates a contribution to
$K_{aA;bB}^{\pm}(E,\omega)$ of 
\begin{widetext}
\ba
\label{eq:K(4beta)}
K^{(\beta)}_{aA;bB}(E,\omega)&=&
\frac{\delta_{ab}\langle H_{BW}H_{UA}\rangle\langle
H^2\rangle_{WU}}{E_+^5 E_-}
+
\frac{\delta_{BA}\langle H_{au}H_{wz}\rangle\langle
H^2\rangle_{uw}}{E_+ E_-^5}
\nonumber \\
&&-2\times\frac{-1}{4}\times\left(
\frac{\langle H_{au}H_{vb}\rangle \langle H_{BA}H_{uv}\rangle}
{E_+^2 E_-^4}+
\frac{
\langle H_{ab}H_{VU}\rangle \langle H_{BV}H_{UA}\rangle}
{E_+^4 E_-^2}
\right) 
\nonumber \\
&&
-\frac{-1}{4}\times
\left( 
\frac{
\langle H_{au}H_{BA}\rangle \langle H^2\rangle_{ub}
}{E_+^2 E_-^4}
+
\frac{
\langle H_{ab}H_{BU}\rangle \langle H^2\rangle_{UA}
}{E_+^4 E_-^2}
+
\frac{
\langle H^2\rangle_{au} \langle H_{ub}H_{BA}\rangle
}{E_+^2 E_-^4}
+\frac{
\langle H_{ab}H_{UA}\rangle \langle H^2\rangle_{BU}}
{E_+^4 E_-^2}
\right)
\nonumber \\
&&
+2\times\frac{1}{6}\times\frac{\langle H^2\rangle_{ab}\langle H^2\rangle_{BA}}{E_+^3 E_-^3}
+4\times\frac{1}{6}\times\frac{\langle H_{au}H_{BU}\rangle \langle H_{ub}H_{UA}\rangle}
{E_+^3 E_-^3}
\ea
\end{widetext}
This equation is depicted graphically in Fig.\ref{fig:Hresolved}.

Three aspects of this exercise are worth noting. First, a 
given \lv diagram contains a partial
sum of several Green function diagrams. Second, the 
Green function diagrams are all
added with positive weights. Third, one's conventional intuition 
about the importance
of diagrams in Green function perturbation theory 
is suspect when carried over to \lv 
perturbation theory. In this example, a 
seemingly simple, non-crossing \lv diagram actually
sums some of the crossing diagrams in Green function perturbation theory.

\subsubsection{The algorithm}
\label{subsec:unintalgo}
To apply the method illustrated above to all diagrams we need to automate
the procedure for keeping track of the topology of disorder contractions.
This can be accomplished by deforming the problem defined by
Eqns.\,(\ref{eq:L}) and (\ref{eq:Pi}) to the form 
\ba
{\hat{\cal L}}(p,h)&=&p\, {\hat H}\circ {\hat 1} - h\, {\hat 1}
\circ {\hat H}  \nonumber
\label{eq:Lph}
\\
{\hat \Pi}(\omega,p,h)&=&[\omega-{\hat{\cal L}}(p,h) + i \delta ]^{-1}.
\label{eq:Piph}
\ea
Here, $p$ and $h$ are arbitrary parameters which
record
whether $H$ acts on a
retarded or an advanced Green function line.
Consider any particular \lv diagram at order $2n$, which we label below
with $\mu$. Its contribution to $\Pi(\omega)$ is
\be
\frac{\langle
\hat L^{2n}\rangle_\mu}{(\omega+i\delta)^{2n+1}} = 
\sum_{m \alpha}A^{
(\mu)}_{m\alpha}{p^{2n-m} h^{m}}\frac{\langle
\hat H^m
\hat H^{2n-m}\rangle_{\alpha \mu}}{(\omega+i\delta)^{2n+1}},
\ee
where $A^{(\mu)}_{m\alpha}$ are combinatorial coefficients and
$\langle\rangle_\mu$ denotes a particular subset of Wick contractions
of $\langle \hat{\cal L}^{2n} \rangle$ corresponding to $\mu$ 
(and similarly for $\langle\rangle_{\alpha\mu}$ in the case of
$\langle \hat{ H}^{2n} \rangle$).
The diagram gives rise to a contribution to $K_{aA;bB}^{\pm}(E,\omega)$ of
\be
\sum_{m\alpha}A^{
(\mu)}_{m\alpha}(-1)^m
\frac{m!(2n-m)!}{(2n)!}\frac{\langle \hat H^m
\hat H^{2n-m}\rangle_{\alpha \mu}}{E_{+}^{m+1}E_{-}^{2n-m+1}} \,.
\ee
This step can be automated
by making the substitutions: $ \omega +i\delta \rightarrow  E_+ E_-$,
$ p  \rightarrow   {x E_+}/{z} $, and
$h  \rightarrow  -{y E_-}/{z} $,
and then, to attach the binomial coefficients, resorting to the identity 
\be
\int_{0}^{\infty}dx\int_{0}^{\infty} dy \oint\frac{dz}{2\pi i z} 
\exp({-x-y+z})\,
\frac{x^m y^n}{z^l}=\frac{m! n!}{l!} \ ,
\ee
where the $z$ integration contour encloses the origin 
once in the anticlockwise direction. 
Thus the algorithm for energy un-integrating the 
Liouvillian perturbation theory is
reformulated as a particular integral transform.

To summarize, $ K (E,\omega)$ can be recovered from $ \Pi (\omega)$ 
by first generalizing $\Pi(\omega)$ to $\Pi(\omega,p,h)$ for a
\lv as in Eq.\,(\ref{eq:Lph}) and  then carrying out the
integrals in
\begin{widetext}
\be
 K (E,\omega)=\int_{0}^{\infty}dx\int_{0}^{\infty} dy
\oint\frac{dz}{2 \pi i z} \exp({-x-y+z})  
\Pi(E_+ E_-,\frac{x}{z}E_+,-\frac{y}{z}E_-)\,.
\label{eq:triple}
\ee
\end{widetext}
As the objects on both sides are formally defined by their diagrammatic series,
this is an exact relationship between them. 
Given a finite set of \lv diagrams, this procedure can clearly be carried out
diagram by diagram.
On the other hand, the procedure may prove too cumbersome to 
deal successfully 
with a particular approximate resummation of Liouvillian perturbation
theory.
It turns out that our program can be carried through 
for the $N_L = \infty$ limit of MSZ, applied to random matrix ensembles, as we
show in Section \ref{RMT}. By contrast, for the quantum Hall problem
even the $N_L = \infty$ limit requires numerical solution 
of an integral equation to obtain the Liouvillian.
The chore of energy unintegration 
in this case is much more involved and we will consider
it only in a special limit. It is conceivable that in
still other cases one may need to resort to numerical resummations of the
unintegrated series.


\section{$1/N_L$ Expansion}
\label{1/N_Lsec}
No exact solution of a problem
using the Liouvillian formalism is
known at present.
It is therefore pertinent to ask whether there are
useful, natural approximations in the \lv approach that are hard to 
uncover in the standard approach. The simplest one, suggested by
the interpretation of the \lv as a random Hamiltonian in its own
right, is the self-consistent Born approximation (SCBA),
employed for this purpose by SMG. In the supersymmetric
functional integral approach of MZS, the SCBA is a saddle point 
approximation, but we will instead take the diagrammatic route, in which
it is a summation of all non-crossing diagrams. In either case,
the procedure is justified formally by deforming the problem to
one with $N_L$ \lv flavors and taking the
$N_L=\infty$ limit. One can then examine the stability of the solution by 
perturbing in $1/N_L$, in the hope that most of the structure in
the problem of interest, $N_L=1$, survives to large \nl.

The recipe for introducing $N_L$ flavors into 
a general Liouvillan problem is as follows. 
One replicates the original $N^2$ dimensional 
space of bilinears $N_L$ times and allows for scattering between different 
flavors. Then
\ba
\Pi^{ij}_{aA;bB}&=&( a,A; i|\frac{1}{\omega-\hat{\cal L} + i \delta}| 
b,B;j )\,,  
\ea
with
\ba
\langle
\hat{\cal L}^{kl}\hat{\cal L}^{mn}
\rangle&=&\frac{1}{N_L}
(\delta_{kn}\delta_{lm}+\delta_{km}\delta_{ln})
\langle \hat{\cal L}^{kl}\hat{\cal L}^{kl} \rangle.
\label{eq:LN}
\ea
where in the Liouvillian correlator we have indicated only the flavor
indices.
The replicated Liouvillian can be thought of as an $N_L \times N_L$ 
matrix 
with individual elements which are themselves
$N^2 \times N^2$ Liouvillians of the original, 
single 
flavor, problem constructed from different realizations of disorder. Moreover, 
$\hat{\cal L}^{kl}=\hat{\cal L}^{lk}$, so that $({N_L^2+N_L})/{2}$ 
independent realizations 
of disorder are necessary to construct a single realization of the replicated 
problem. The entire $N_L N^2 \times N_L N^2$ matrix is Hermitian.  
The construction we have described is, in the terminology of MSZ,
the orthogonal  generalization 
of the $N_L=1$ 
problem and is the one they have argued is useful in generating a 
$1/N_L$ expansion for 
the localization length exponent for the quantum Hall plateau transition.

To see how the expansion goes, consider as $N_L\rightarrow \infty$ the  
low order diagrams shown in Fig.\,\ref{fig:largen}. 
\begin{figure}
\label{fig:largeN}
\epsfxsize=2.5truein
\centerline{\epsffile{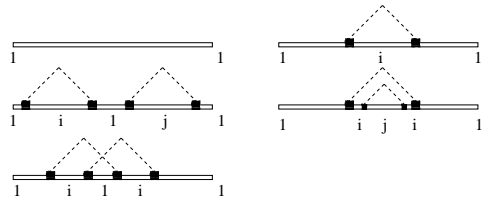}}
\caption{Liouvillian diagrams up to fourth order in ${\hat{\cal L}}$.
The last diagram contains a crossing  and is smaller by a factor $1/N_L$
than the others, which are of order $N_L^0$. 
Only flavor indices are shown explicitly.
}
\label{fig:largen}
\end{figure}
Evidently, diagrams with crossed disorder 
lines are suppressed by factors of $1/N_L$
relative to those without crossed lines.
This suppression of crossed diagrams continues in higher 
orders of the expansion. Thus the leading approximation at large $N_L$
is to
sum all non-crossing terms, which are of order $N_L^0$.
As usual, this sum can be carried out by solving the equivalent
self-consistency 
equation 
\be
\Pi^{ij}_{aA;bB}=\frac{\delta_{ab}\delta_{AB}\delta_{ij}}{\omega + i
\delta}+\frac{1}{\omega + i \delta} 
\left[\langle \hat{\cal L} \cdot \hat{\Pi} \cdot \hat{\cal L}\rangle \cdot
\hat{\Pi}\right]^{ij}_{aA;bB}
\label{eq:SCBA}
\ee
represented graphically in Fig.\,\ref{fig:SCBA}.
\begin{figure}
\epsfxsize=2.5truein
\centerline{\epsffile{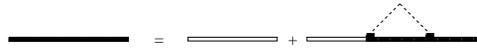}}
\caption{The Liouvillian SCBA in diagrammatic form.
Clear lines represent bare propagators, and
filled ones the full propagator, $\hat \Pi(\omega)$.
}
\label{fig:SCBA}
\end{figure}
At next order ($N_L^{-1}$)  one needs to sum  maximally 
crossed graphs and rainbow diagrams shown in Fig. \ref{fig:mc}.
At any order the Liouvillian Green function 
is diagonal in the flavor index
\be
\Pi^{ij}_{aA;bB}(\omega)=\delta_{ij}\Pi_{aA;bB}(\omega)
\ee
and $\Pi_{aA;bB}(\omega)$ is taken as an approximation to the $N_L=1$
Liouvillian Green function.


\section{Random Matrix Theory}
\label{RMT}

We now turn to the problem of random matrix statistics as a test
of the \lv large $N_L$ approach. The large number of known exact
results on this problem
and the relative simplicity of calculations
make it an ideal test case. While there
is no phase
transition (all Green functions are analytic inside
the band), we will see
that there is sufficient non-trivial structure in the statistics
of the eigenvalues
that can be used diagnostically. 
Oddly enough, we will find that the \lv analysis will yield
some Green functions that {\it are} singular inside the band (see
Section IVE).

The program is this: for 
$N \times N$ random matrices, we introduce an enlarged
\lv with $N_L$ flavors. Now we have (at least) two different 
limits to consider. The standard one takes $N \gg
1 $ at $N_L=1$, and leads to a familiar set of simplifications
such the Wigner semicircle law for the density of eigenvalues
which are summarized below. It also leads to a perturbation theory in
$1/N$  which is the zero dimensional version of impurity 
averaged perturbation theory for single-particle Green 
functions in disordered conductors. 
The new limit we will consider is $N_L \gg 1$
at fixed $N$. As in all large-$n$ 
expansions, this will prove useful
if the large $N_L$  problem is sufficiently smoothly connected to 
the $N_L=1$ problem of interest. While we will find it useful
to then take $N \gg1$ (but keeping $N \ll N_L$) as well, the possibility
of finding non-trivial information on finite matrices is
a potential asset of this limit. 
Similarly, in the application to the quantum Hall plateau transition, 
the infinite volume limit
is taken after $N_L\rightarrow \infty$.

Before proceeding we collect some standard definitions and
results on the random matrix theory for ease of comparison with the
following \lv analysis.

\subsection{Ensembles}
We will be interested in random Hermitean matrices 
$\hat{H}$ drawn from
the Gaussian unitary ensemble (GUE)
if $\hat H$ is complex, and the Gaussian orthogonal
ensemble (GOE) if $\hat{H}$ is
required to be purely real. 
For these ensembles it is sufficient to
specifiy  the correlators of the matrix elements, 
\ba
\langle H_{ab}H_{cd}\rangle &=&\frac{v^2}{N}(\delta_{ad}\delta_{bc}+
\alpha\delta_{ac}\delta_{bd})\,,
\label{eq:HH}
\ea 
where $\alpha =0$ for GUE and $\alpha =1$ for GOE.
The variance has been normalized by the matrix size, $N$, 
to produce a  
bandwidth which remains finite as $N\rightarrow\infty$. 

\subsection{Correlators}

In addition to the fundamental one and two particle Green functions defined
in Section II, we will be interested in the 
the density of states (DOS) of $\hat H$
\be
\rho(E)= \sum_n \delta(E-\epsilon_n)=-\frac{1}{\pi}{\rm Im}{\rm
Tr}\left[\frac{1}{E-\hat{H}+i \delta}\right] 
\ee
and its disorder averaged correlation function 
\be
R(E,\omega)=
\langle{\rho(E+\frac{\omega}{2})\rho(E-\frac{\omega}{2})}\rangle \ .
\ee
$R(E,\omega)$ is related to the two-particle Green functions $K$ by
\be
R(E,\omega)=\frac{1}{4\pi^2}
\sum_{aA}[K^{+-}_{aA;aA}+K^{-+}_{aA;aA}-K^{++}_{aA;aA}
-K^{--}_{aA;aA }] \ .
\ee
The \lv Green function allows us to extract the Liouvillian density of 
states (LDOS)
\be
\Omega(\omega)=
\sum_{mn} \delta(\omega-\lambda_{mn})
= -\frac{1}{\pi}{\rm Im} 
{\rm Tr}\left[ \frac{1}{\omega- \hat{\cal{L}}+i \delta}\right]
\ee
and it follows from 
Eq.\,(\ref{eq:lee})  that 
$\Omega(\omega)=\int dE R(E,\omega)$. 

Disorder averaging (at any $N$) simplifies the structure of various 
correlators
\begin{widetext}
\ba
\langle{G}_{mn}(E)\rangle&=&\delta_{mn} G(E) \nonumber \\
\langle K_{aA;bB}(E,\omega)\rangle
&=&\delta_{ab}\delta_{AB}K_{1}(E,\omega)+
\delta_{aA}\delta_{bB}K_2(E,\omega)+\delta_{aB}\delta_{Ab}K_3(E,
\omega)\nonumber \\
\langle \Pi_{aA;bB}(\omega)\rangle&=&\delta_{ab}\delta_{AB}\Pi_{1}(\omega)+\delta_{aA}
\delta_{bB}\Pi_{2}(\omega)+\delta_{aB}\delta_{Ab}\Pi_{3}(\omega) \ 
\label{eq:avgdPi}
\ea
\end{widetext}
with the constraints $K_2=K_3$, $\Pi_2=\Pi_3$ for GOE and
$K_3=\Pi_3=0$ for GUE. 
Finally,  the \lv density of states  has the form
\ba
\Omega(\omega)
&=& - \frac{N^2}{\pi}\left({\rm Im} \Pi_1(\omega) +\frac{{\rm Im}\Pi_2(\omega)+
{\rm Im} \Pi_3(\omega)}{N}\right)
\ ,
\label{eq:Omega}
\ea
where the powers of $N$ arise from taking a trace over the index
structure in Eq.\,(\ref{eq:avgdPi}).

\subsection{Standard results: large N at $N_L=1$}

At $N=\infty$ the SCBA for the single-particle Green function,\cite{foot:sqrt}
\be
G_{mn}^\infty(E)=
\delta_{mn} \frac{E}{2v^2}\left(1-\sqrt{1-{4v^2}/{E^2}}
\right)
\label{eq:Wigner}
\ee
is exact, and hence the DOS is the Wigner semicircle: 
$\lim_{N\rightarrow \infty} \langle \rho
(E)\rangle /N=({1}/{\pi v})\sqrt{1-({E}/{2v})^2}$. 
For $N$ large but finite, the deviations from this
limiting form are small. For example, the leading behavior
of the DOS at energies above the large $N$ upper band edge, $E=2v$, is 
$\rho(E) \sim \exp(-4N(E/v - 2)^{3/2}/3)$.
Higher order correlators of the DOS, including $R(E,\omega)$,
can be discussed in each of two limits. 
First, taking $N\to \infty$ with energy arguments fixed and all
different, they factorize. For example, with $\omega \not= 0$,
\be
R_{\infty}(E,\omega)
=
\langle 
\rho_{\infty} (E+\frac{\omega}{2})
\rangle 
\ \langle \rho_{\infty} (E-\frac{\omega}{2})\rangle .
\ee
Second, by scaling the separation of energy arguments with
the mean level spacing, universal corrleation functions 
are obtained, which are dominated at small energy separations
by level repulsion. Thus, for $N\omega \ll 1$,
%
%
\be
R(E,\omega )\propto  (\omega N)^\beta \rho(E)^2\ \,,
\label{beta}
\ee
with $\beta=1$ for GOE and $\beta=2$ for GUE.

Combining these with our earlier listing of the general properties
of \lvs we conclude that the exact LDOS 
will have almost all its weight within the range $-4v < \omega< 4v$
for $N \gg 1$. For $N=\infty$ it is exactly zero outside and
vanishes with zero slope (quadratically
$\sim(4v\pm\omega)^2$) near the edges. Finally, at finite $N$,
there should be a
pseudogap of width $v/N$ near $\omega=0$ with details depending on the  
symmetry of the ensemble.

\subsection{RMT \lv
}
\subsubsection{$N_L=\infty$, $N$ arbitrary}
For the GUE Eq.\,(\ref{eq:SCBA})  becomes 
\ba
\Pi_1(\omega)&=&\frac{1}{\omega+i \delta}+
\frac{2 v^2 \Pi_1(\omega)^2}{\omega+i\delta}
\nonumber\\
\Pi_2(\omega)&=&-\frac{2v^2}{N (\omega+i \delta) } \Pi_1(\omega)^2 =
\frac{1}{N}\left(\frac{1}{\omega+i \delta}-\Pi_1(\omega)\right)
\nonumber\\ 
\Pi_3(\omega)&=&0
\label{GUEsc}
\ea
The solution is similar to that for
$N=\infty$ RMT (cf. Eq. \ref{eq:Wigner}). Using Eq.\,(\ref{eq:Omega}) the LDOS is
\ba
\Omega(\omega)
&=&  \delta(\omega)+\frac{N^2-1}{\pi\sqrt{2v^2}}
\sqrt{1-\frac{\omega^2}{8v^2}}
\label{eq:GUELDOS}
\ea
for $|\omega|<2 \sqrt{2} v$ and zero otherwise. 
Notice that the support of the spectrum at $N_L=\infty$ is completely
independent of $N$. 

Turning to GOE, the self-consistency  condition
reduces to the quadratic equations
\begin{widetext}
\ba
\Pi_1(\omega)&=&\frac{1}{\omega+i \delta}+\frac{2 v^2(N+1)}{N(\omega+i
\delta)}( \Pi_1(\omega)^2
-\Pi_3(\omega)^2) \nonumber \\
\Pi_3(\omega)&=&\Pi_2(\omega)=\frac{2v^2}{N
(\omega+i \delta)}(\Pi_3(\omega)^2-\Pi_1(\omega)^2)
=\frac{1}{N+1}\left(\frac{1}{\omega+i \delta}-\Pi_1(\omega)\right)
\ea
with solution
\ba
\Pi_1(\omega)&=&\frac{-4v^2+(1+N)(N (\omega+i \delta)^2-
\sqrt{16 v^2-8N(1+N)v^2
(\omega+i \delta)^2+N^2(\omega+i \delta)^4})}{4N(2+N)v^2(\omega+i \delta)}\,,
\ea
leading to the LDOS
\ba
\Omega(\omega)&=& -\frac{1}{\pi}\left(N^2-\frac{2N}{N+1}\right){\rm Im}
\Pi_1(\omega)-\frac{2N}{\pi(N+1)}{\rm Im}\frac{1}{\omega+i \delta}\,.
\label{LDOS-GOE}
\ea
\end{widetext}
As $N\rightarrow \infty$ the leading, ${\cal O}(N^2)$, piece of this
result is identical to the corresponding piece of $\Omega(\omega)$ for the GUE.
For finite N, however, the LDOS is
zero in the window $|\omega|< \sqrt{2v^2}/N$, in stark contrast with the
GUE result (or anything else one might expect of a disordered system:
according to this approximation the spectrum of the disordered problem
is {\it completely gapped}).
These results are illustrated in Fig. \ref{LDOS}.
\begin{figure}[h]
\epsfxsize=2.5truein
\centerline{\epsffile{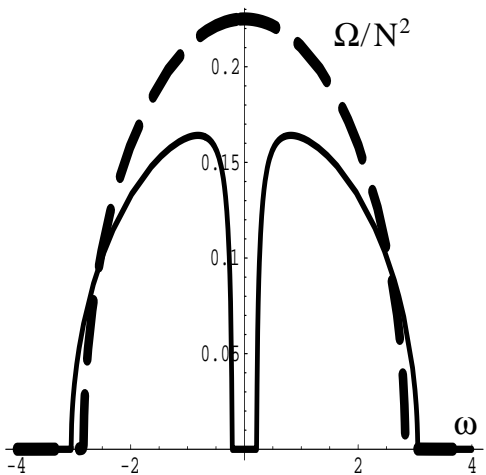}}
\caption{LDOS scaled by $N^2$ at $N_L=\infty$ with disorder strength $v=1$. 
Solid line is GOE (Eq. \ref{LDOS-GOE}) at $N=6$ while the broken one is GOE (Eq. \ref{LDOS-GOE}) at $N=\infty$
and GUE (Eq. \ref{eq:GUELDOS}) at arbitrary $N$. The delta function at zero frequency has been suppressed.
}
\label{LDOS}
\end{figure}

There are at least four important differences between these
exact $N_L=\infty$ results and the corresponding, correct, $N_L=1$ solutions:
\begin{itemize}
\item
At $N_L =\infty$ results for both GUE and GOE imply a bandwidth of
$2\sqrt{2}v$ for the underlying random matrices. 
This is a factor of $\sqrt{2}$ less than that of the
correct solution. 
\item 
The infinite gradient  of the $N_L=\infty$ expression for $\Omega(\omega)$ 
at the \lv band edges is at odds with the exact result, which
has vanishing slope. Contributions to the LDOS at the
\lv band edges come from
pairs of random matrix eigenvalues which are near opposite 
random matrix band edges.
If one assumes that $R$ will factorize in such cases, which is
certainly true of the exact result and should be expected anyway
for such widely separated eigenvalues,  the form found for LDOS
implies a DOS {\it divergent} near the random matrix 
band edges, with  $\rho(E)\sim (v \sqrt{2}\pm E)^{-1/4}$.
In fact, we shall see that $R$ does not factorize at $N_L=\infty$,
which turns out to be the central pathology of that
limit.

\item
Although there is a delta function at zero frequency,  
arising from zero eigenvalues of $\hat{\cal L}$ (see Eq.\,(\ref{eq:lee})), 
its 
weight is $1$, and not $N$ as it should be.
\item Finally, the two solutions are qualitatively 
wrong in the small frequency 
limit ($|\omega|\lesssim 1/N$) in opposite ways. The GUE LDOS is 
non-zero with no indication of level repulsion, while the GOE
develops a {\it clean} gap over a region of ${\cal O}(1/N)$.
\end{itemize}
The $N_L=\infty$ result above is clearly not an adequate approximation 
to $N_L=1$ even for the purposes of computing properties of the
\lv. In the next section we will see that matters get worse when
we energy unintegrate the $N_L=\infty$ result. Before turning to that,
we first ask whether one may construct a smooth interpolation between the two 
limits by perturbing in $1/N_L$. We offer evidence that this is
unlikely, both
from considering the $1/N_L$ expansion analytically,
and from studying the problem numerically
for a range of $N_L$ values.

\subsubsection{Leading $1/{N_L}$ corrections}
As discussed in Section \ref{1/N_Lsec}, the
leading correction for large $N_L$ to the $N_L = \infty$ 
Liouvillian Green function 
is given by the maximally crossed diagrams. These are
resummed by two consecutive geometric series, represented graphically
in Fig. \ref{fig:mc}.
\begin{figure}[h]
\epsfxsize=2.5truein
\centerline{\epsffile{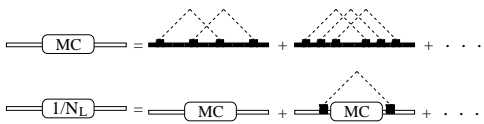}}
\caption{All diagrams for ${\hat \Pi}(\omega)$ at ${\cal O}(1/N_L)$.
}
\label{fig:mc}
\end{figure}

Denoting the $N_L=\infty$
solution by $\Pi_1$,
the result of this calculation for the GUE is 
\ba
\delta \Omega(\omega)&=&\frac{1-N^2}{N_L \pi}
{\rm Im}(\frac{2(\Pi_1)^5 v^4}{(1-v^2 (\Pi_1)^2)(1-2v^2(\Pi_1)^2)})
\nonumber
\ea
\begin{figure}[h]
\epsfxsize=2.5truein
\centerline{\epsffile{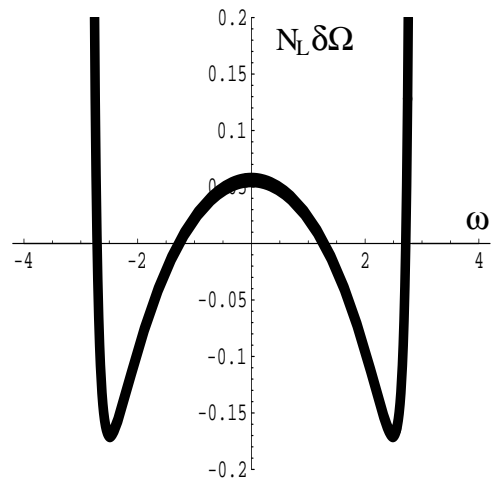}}
\caption{$1/N_L$ correction to the LDOS for GUE
}
\label{deltaOmGUE}
\end{figure}
As is the case with $1/N$ corrections to the RMT DOS,
the $1/N_L$ correction to the LDOS is finite near
$\omega=0$ and develops a singularity at band edges,
as illustrated in Fig. \ref{deltaOmGUE}. In the case of
the GOE, the $1/N_L$ correction, which for brevity we do not display, 
is divergent at all four band edges.
These divergences show that any attempt to fix the bandwidth
problems of the $N_L = \infty$ limit, if feasible, must involve 
an analysis including contributions at all orders in $1/N_L$.
We have not pursued this analysis further. Instead, we
offer evidence from numerical studies that expansion about
$N_L=\infty$ is unlikely to yield useful information on
behavior at $N_L=1$.

\subsubsection{Numerics at small $N_L$}
\label{sec:RMTnumerics}
The simplicity of the random matrix problem affords us a different, more
direct line of attack via exact diagonalization. Specifically, we have
diagonalised numerically \lv RMT for a range of $N_L$ and $N$, both to
test the relevance of  analytic results above and also to search for
new features, especially in the interesting regime of small $N_L$.
Representative plots of $\Omega(\omega)$ are shown for the GUE
in Fig.\,\ref{fig:GUELDOS} and for the GOE
in Fig.\,\ref{fig:GOELDOS}, both covering a range in $N_L$ at $N=2$.
We have searched for, but found no significant qualitative differences at
larger N. 
Our conclusions from these figures are as follows.
First, 
the gross of features of the $N_L=\infty$  limit (see Fig. \ref{LDOS}) 
are
already observable at $N_L$ as  small as $N_L=8$.
Second,
\lv theories at $N_L > 1$ 
are sufficiently different from the theory at $N_L=1$,
even for $N_L=2\ {\rm and}\ 3$, that the quantitative
utility of any $1/N_L$ corrections seems questionable.
Readers should note especially the rapid change in
the distribution of small eigenvalues as $N_L$ is increased.
Third, while one clearly cannot hope to investigate analyticity
conclusively by these means,  
there are two features to the numerical results
which suggest behavior non-analytic in $1/N_L$ at $N_L=\infty$.
One of these is the number of states outside the $N_L=\infty$ band,
which appears to decrease
exponentially in $N_L$, as in the  band tails of the RMT DOS.
The other is the presence of oscillations in the LDOS for the GUE
at small energies, as a function of $N_L$: $\Omega(0)$ 
vanishes for odd $N_L$ and is non-zero for even ones.
\begin{widetext}

\begin{figure}[h]
\epsfxsize=2.5truein
\epsfig{file=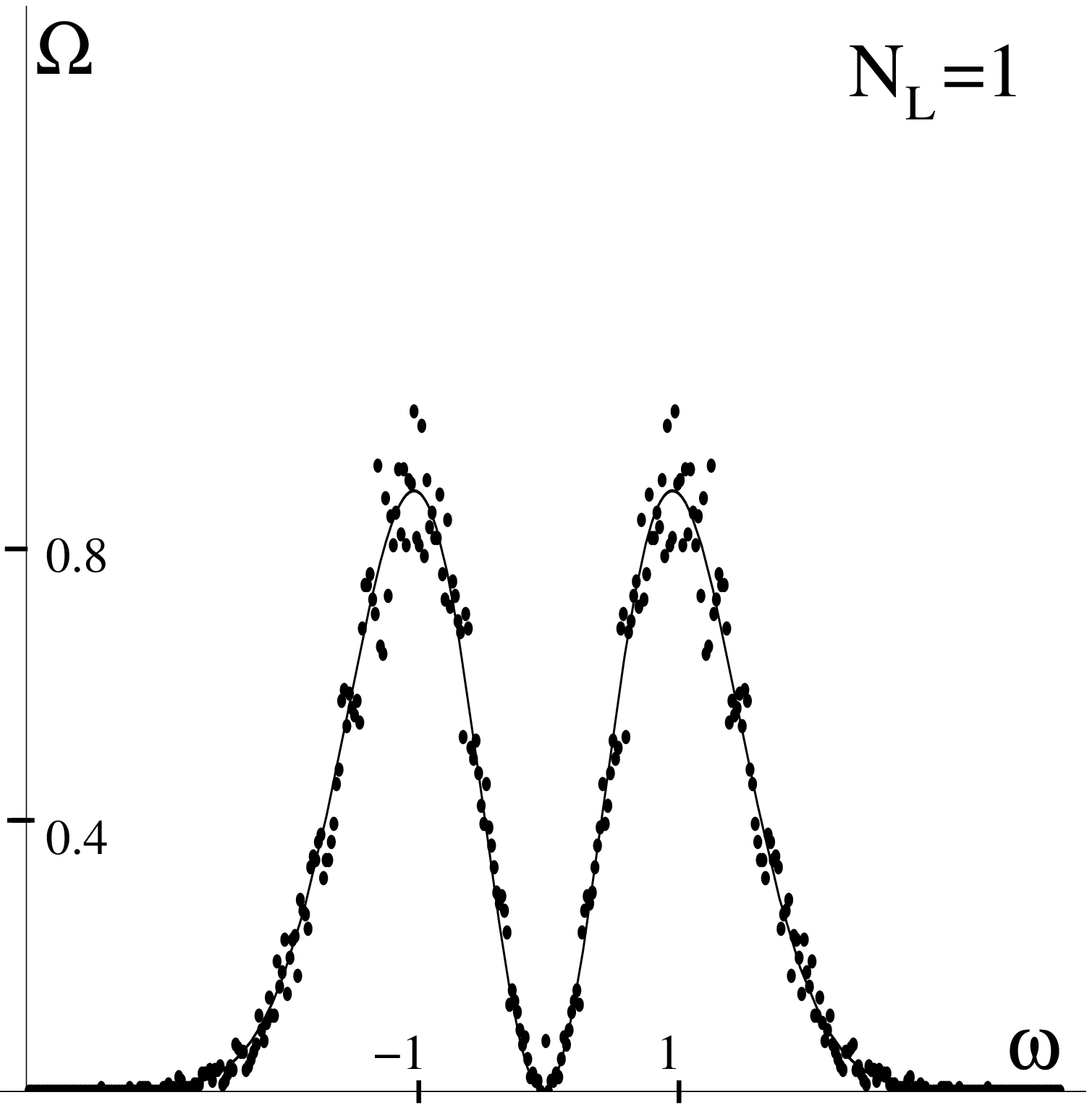,height=1.6in}\epsfig{file=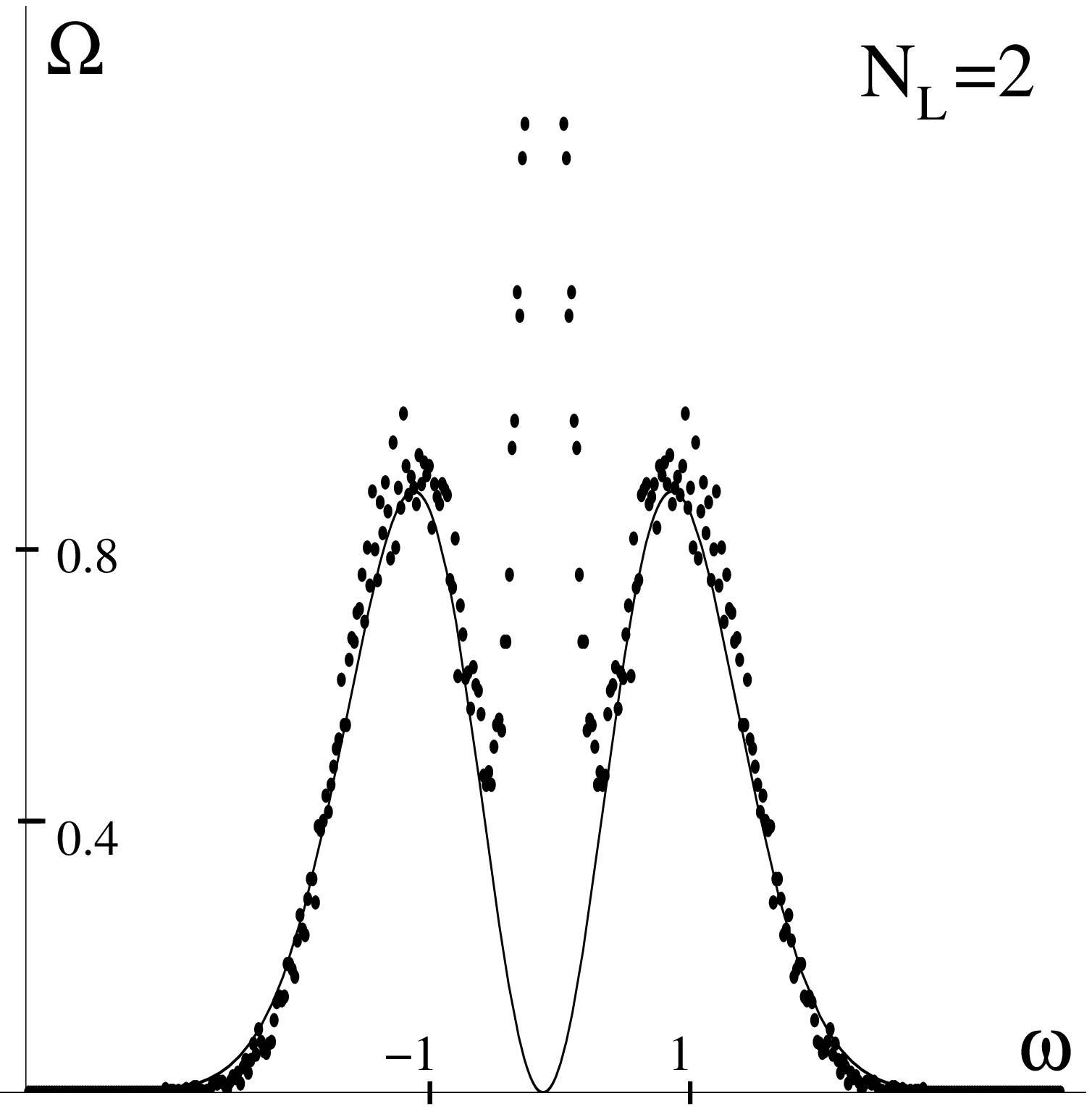,height=1.6in}
\epsfig{file=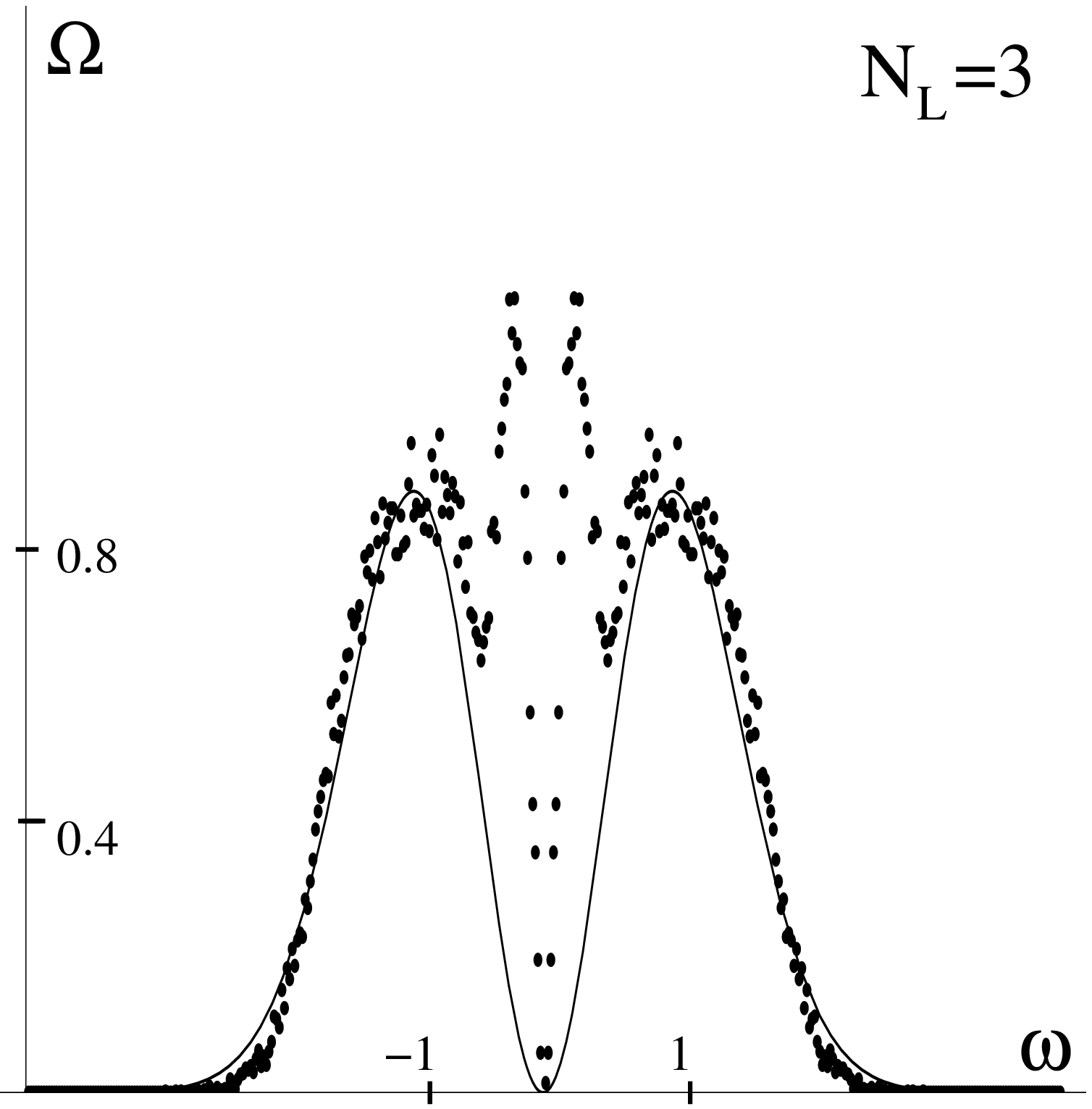,height=1.6in}\epsfig{file=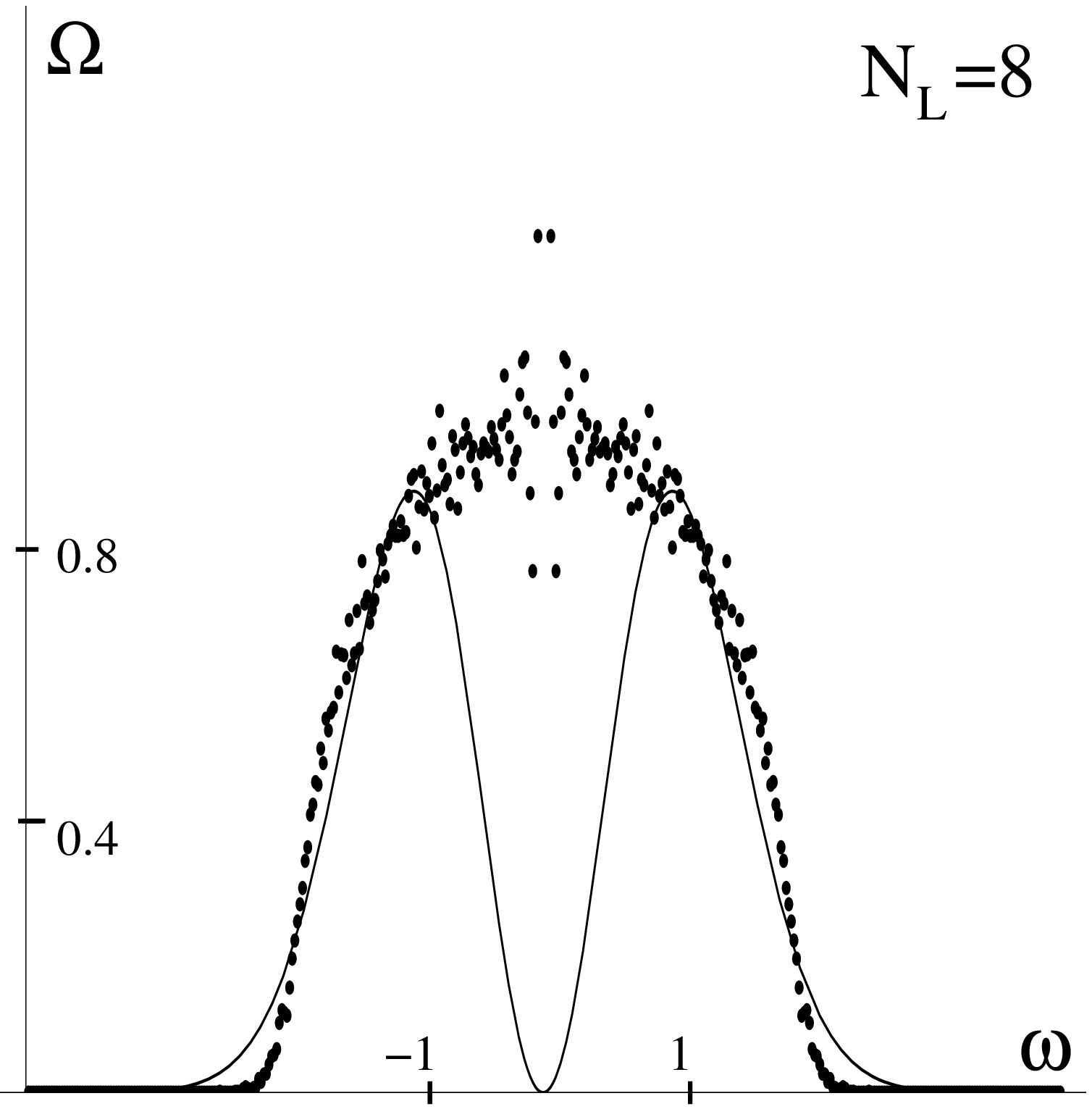,height=1.6in}
\caption{Evolution of $\Omega(\omega)$ from $N_L=1$ to $N_L=8$ for N=2 GUE of random matrices. Solid line is the exact solution for $N_L=1$\cite{exact2}.
Delta functions at zero frequency have been suppressed.}
\label{fig:GUELDOS}
\end{figure}

\begin{figure}[h]
\epsfxsize=2.5truein
\epsfig{file=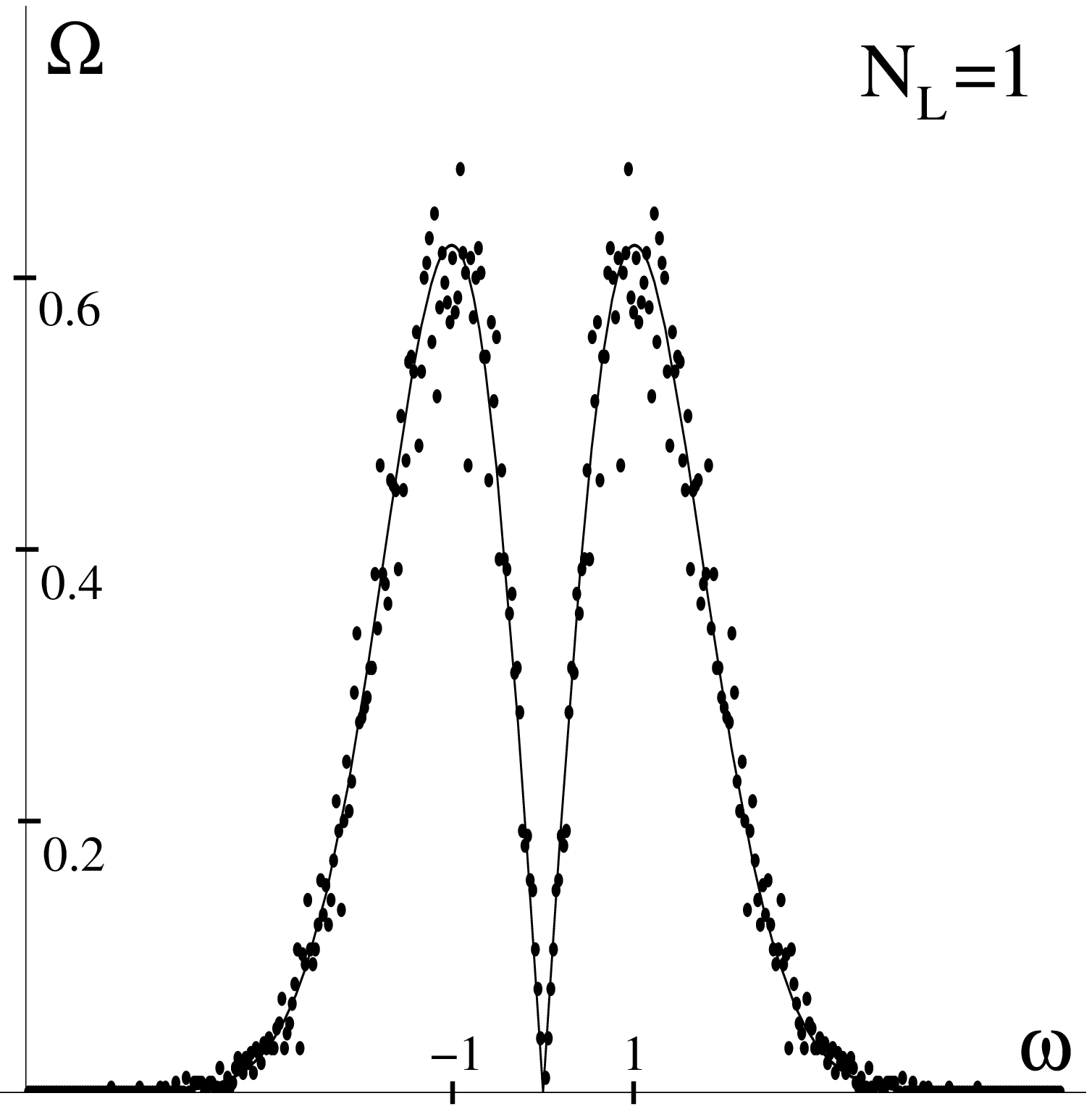,height=1.6in}
\epsfig{file=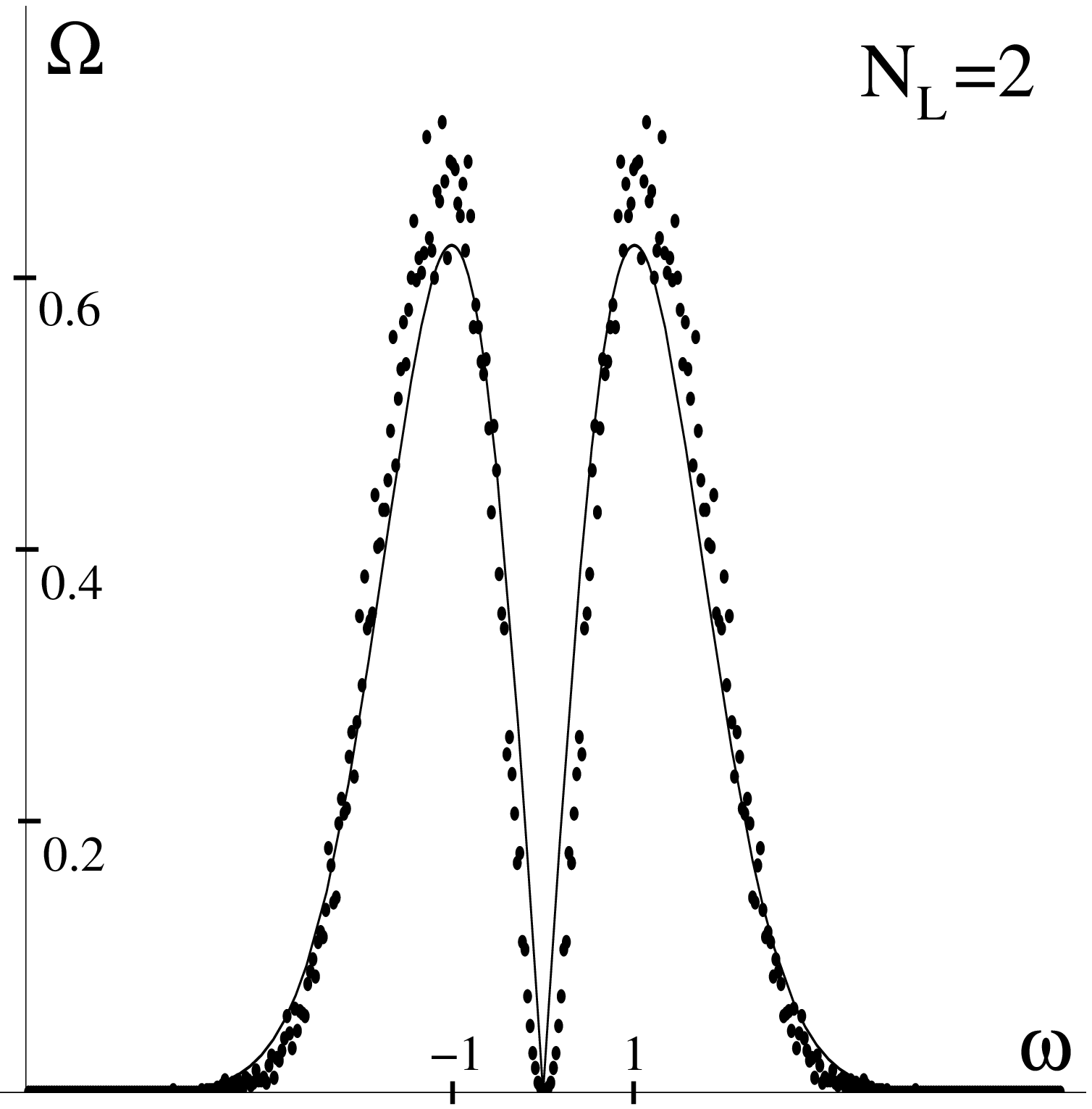,height=1.6in}
\epsfig{file=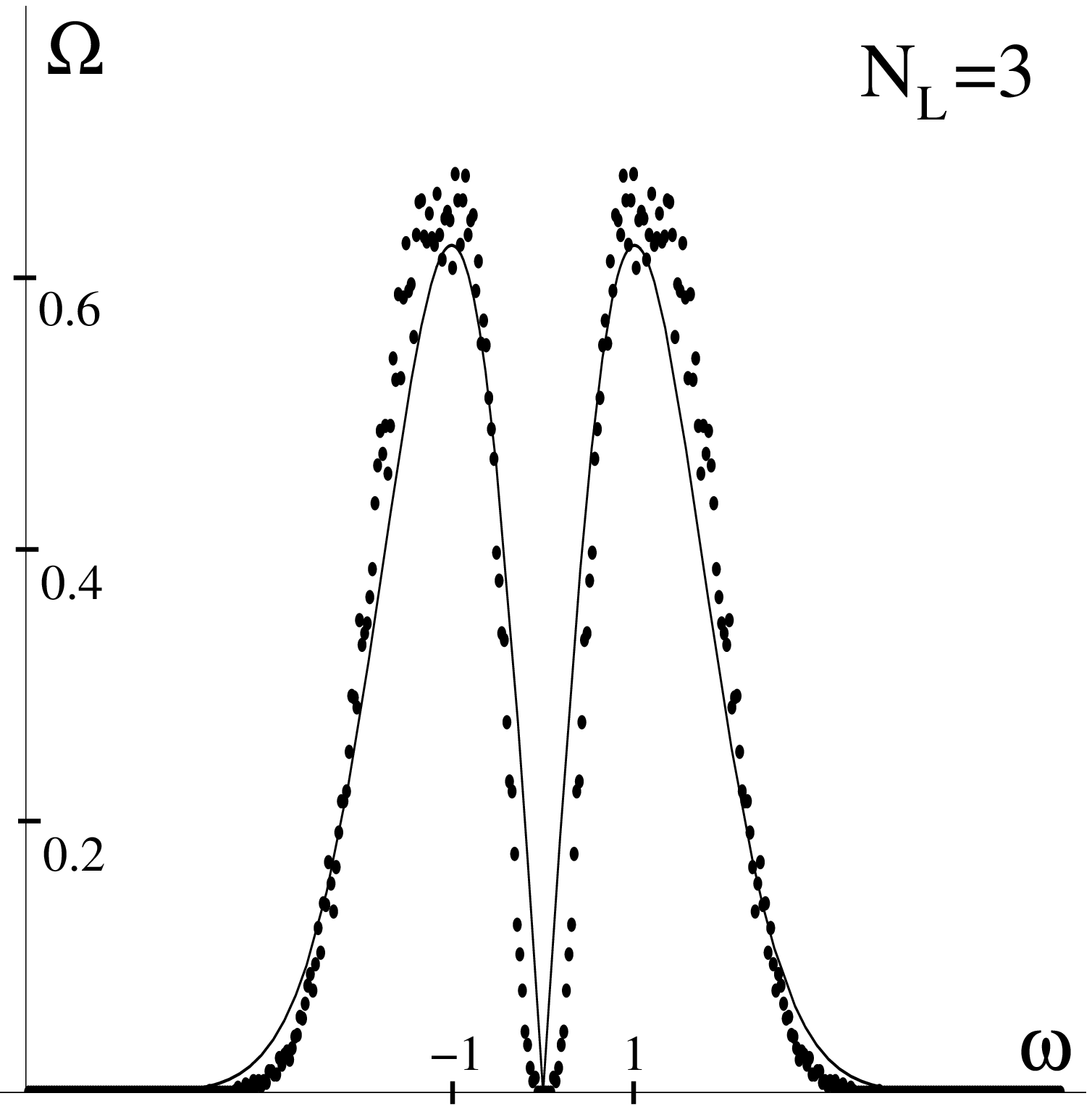,height=1.6in}
\epsfig{file=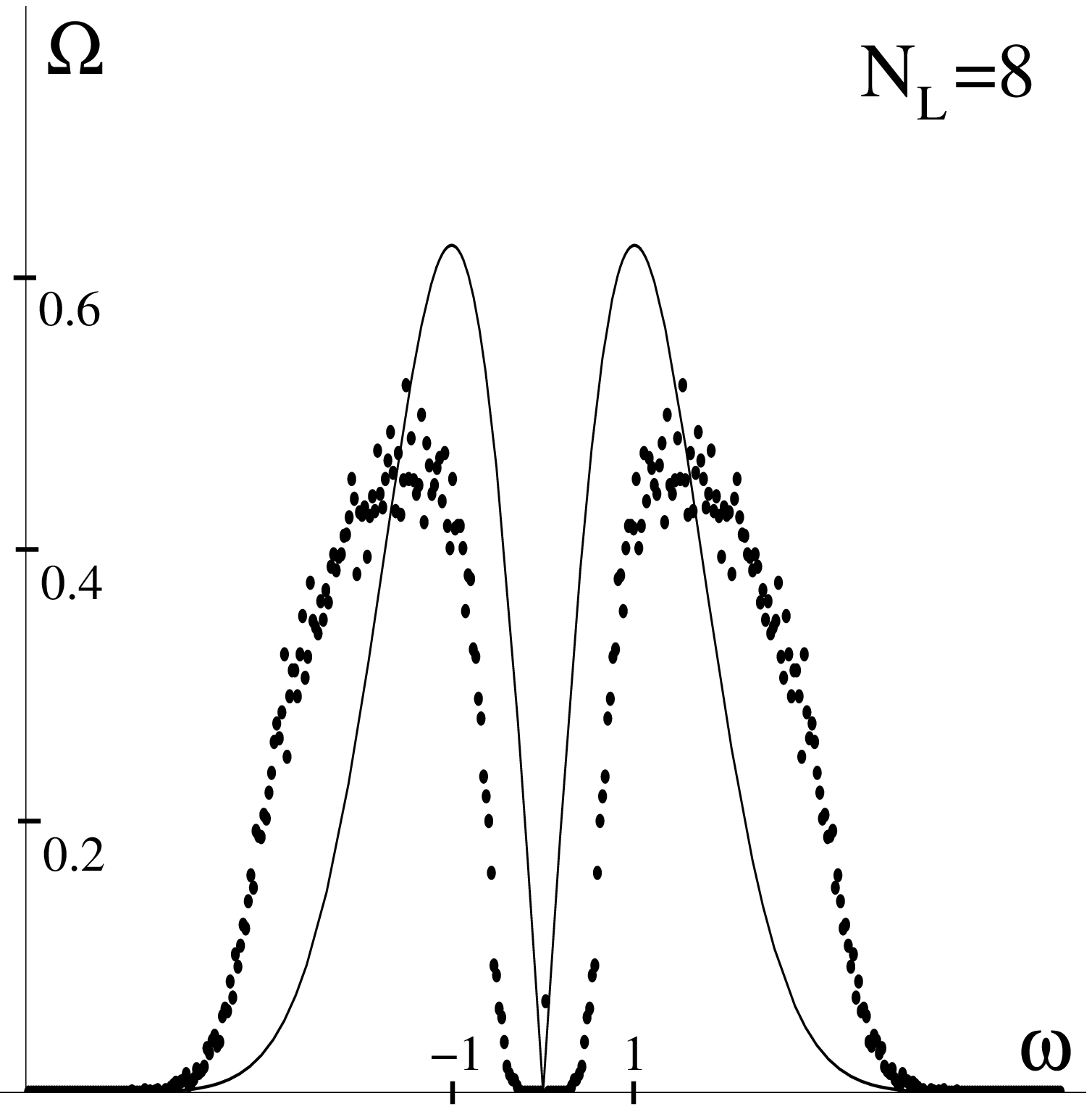,height=1.6in}
\caption{Evolution of $\Omega(\omega)$ from $N_L=1$ to $N_L=8$ for N=2 GOE of random matrices. Solid line is the exact solution for $N_L=1$\cite{exact2}.
 Delta functions at zero frequency have been suppressed.
}
\label{fig:GOELDOS}
\end{figure}

\subsection{Undoing the energy integration}
In this subsection we apply the formalism developed in
Sec IIE to translate the $N_L=\infty$ RMT calculation
into Green function language. We start from Eq.\,(\ref{eq:Piph}),
applied to RMT. 
Summing the non-crossed diagrams with $\hat{\cal L}(p,h)$ leads for
the GUE, to the self-consistency equations, modified from Eq.\,(\ref{GUEsc}) to
\ba
\Pi_1(\omega)&=&\frac{1}{\omega+i \delta}+\frac{\Pi_1(\omega)
v^2}{\omega+i\delta}\left[(p^2+h^2)\Pi_1(\omega)+\frac{(p-h)^2}{N} 
\Pi_2(\omega) \right] \nonumber \\
\Pi_2(\omega)&=&-\frac{2phv^2 \Pi_1(\omega)^2}{N(\omega +i \delta)}+
\frac{v^2 (p-h)^2
\Pi_2(\omega)}{\omega+i \delta
}\left[\Pi_1(\omega)+\frac{\Pi_2(\omega)}{N} \right]\nonumber\\
\Pi_3(\omega)&=&0.
\ea 
\end{widetext}
We now take the further limit $N\rightarrow \infty$
whereon the solution is 
\ba
\Pi_1(\omega)&=&\frac{\omega}{2v^2(p^2+h^2)}\left(1-\sqrt{1-\frac{4(p^2+h^2)v^2}{(\omega+i\delta)^2}}\right) \nonumber
\ea
In this limit the solution for the GOE is identical.

Using Eq.\,(\ref{eq:triple}) we obtain $K^{+-}_{aA;bB}(E,\omega) = 
\delta_{ab} \delta_{AB}
K_1(E,\omega)$ where 
\begin{widetext}
\ba
K_1(E,\omega)&=&
\int_0^\infty\!\!\!\!\! dx \int_0^\infty \!\!\!\!\! dy 
\exp({-x-y})\oint \frac{dz\, e^z\, E_+^{-1} E_-^{-1}}{2\pi i
2v^2[(x E_-^{-1})^2+(y E_+^{-1})^2)]} 
\left[z-\sqrt{z^2-4v^2[(x E_+^{-1})^2+(y E_-^{-1})^2]}\right].
\label{eq:unint}
\ea
We carry out the integral for large positive $E_{\pm}$ 
(see Appendix B for details) and obtain results for general 
$E_{\pm}$ by analytic continuation. We find
\ba
K_1(E_+,E_-)&=&
\frac{1}{2v^2}\left[\sin^{-1}\left(\frac{E_+ \sqrt{E_+^2-4v^2}+E_-
\sqrt{E_-^2-4v^2}}{E_{+}^2+E_{-}^2-4v^2}\right)
-\frac{\pi}{2}\right]
\label{eq:Kunint}
\ea
where the signs of the square roots are positive for real $E_{\pm}>2v$,
and branch cuts in the complex $E_+$ and $E_-$ planes run on the real axes,
from $E_\pm=-2v$ to $E_\pm=2v$.

The very first thing to note about this result is that, unlike the
exact answer 
\ba
K^{+-}_{aA;bB}& =& G^+(E_+) G^-(E-)
= \delta_{ab} \delta_{AB} {E_+ E_- \over 4 v^4} 
\left(1 - \sqrt{1 - {4 v^2 \over E_+^2}} \right)
\left(1 - \sqrt{1 - {4 v^2 \over E_-^2}} \right)\,,
\label{eq:GG}
\ea
\end{widetext}
it does not factor in its dependence upon $E_\pm$. At $N=\infty$, where
the energies $E_\pm$ involve eigenvalues which are at infinite separation
on the scale of the level spacing, this
is manifestly wrong.

To proceed further in this discussion, we need to examine the analytic
structure of Eq.\,(\ref{eq:Kunint}).  Observe that the result is an 
analytic function of real $E_+$ and $E_-$ when 
both $E_+>2$ and $E_->2$. In this region it is the sum of the convergent 
perturbative expansion in inverse powers of $E_\pm$ that is generated
by unintegrating the $N_L = \infty$ series term by term. Also, in
this region $K^{+-}$ is purely real and the choice of the sign of
the imaginary infinitesimals in $E_\pm$ is immaterial: it equally
well describes an approximation to $K^{++}$. 

The series diverges on approaching
the boundary of this region. In the two-dimensional plane
of real $E_+,E_-$, this happens
at the outer boundary of the cross-shape
formed by the four lines $E_+=\pm 2v$ and $E_-=\pm 2v$, shown in
Fig.\,\ref{fig:thecross}.
Inside this cross-shape,
the choice of infinitesimal imaginary parts to
$E_\pm$ is essential to
specify which side one is of the branch cuts in Eq.\,(\ref{eq:Kunint}).
Depending on how this is done, we obtain inside the cross both $K^{+-}$ and 
$K^{++}$, which are no longer equal but are both
generically complex. The development of imaginary parts to the Green
functions occurs when one of their arguments enters the band.
Indeed, in the exact expression, Eq.\,(\ref{eq:GG}), this is trivially true.

Random matrices do not exhibit a phase transition and so we should
expect that $K^{+-}$ and $K^{++}$ are
analytic functions of $E_+$ and $E_-$ except at the band edges,
$E_\pm^2=4v^2$.
This is true of the exact result, Eq.\,(\ref{eq:GG}).
Surprisingly, it is {\it not} true
of the 
correlators derived from Eq.\,(\ref{eq:Kunint}). Specifically,
inside the cross, while
$K^{+-}$ is analytic, $K^{++}$ exhibits singularities on the
circle $E_+^2 + E_-^2 = 4 v^2$. For example, along
the line $E_+ = E_-$, i.e. $\omega =0$, we find
\be
K^{+-}(E, \omega=0) = {\pi \over 4 v^2}
\ee
while
\begin{widetext}
\be
{\rm Re} K^{++}(E, \omega=0) =  
{\pi \over 4 v^2} {\rm sgn} ({E \over v} - \sqrt{2})
\ee
for $0< {E / v} < 2$ and
\be
{\rm Im} K^{++}(E, \omega=0)  = \left\{
\begin{array}{ccc}
\frac{1}{v^2}{\cosh}^{-1}\left( {{E / v} \over 
\sqrt{2 ({E / v})^2 -4}}\right) &{\rm when}& 
\sqrt{2} < {E / v} < 2 \nonumber \\
 - \frac{1}{v^2} \sinh^{-1} \left( {{E / v} \over 
\sqrt{4 - 2 ({E/ v})^2 }}\right) &{\rm when}& \  0 < {E / v} <  
\sqrt{2} \end{array}\right.
\ee
\end{widetext}
where the negative branch is picked for the $  \cosh^{-1}$. These results
are plotted in Fig.\,\ref{fig:ReK++} and Fig.\,\ref{fig:ImK++} 
along with the exact 
Green functions for comparison.

Finally, one can extract from the knowledge of both $K^{+-}$ and $K^{++}$
the correlator of the density of states, which has the simple form
\be
R(E,\omega)=\frac{1}{4\pi v^2}\Theta(4v^2-2 E^2-\frac{\omega^2}{2}) \ 
\ee
It is constant inside the circle $E_+^2+E_-^2=4v^2$ and zero outside.
This solves one remaining puzzle. While the \lv derived two particle
Green functions imply the same RMT band as the the exact ones, in
the sense that both are analytic outside the cross, we noted
previously that the \lv bandwidth was wrong by a factor of $\sqrt{2}$.
We can now recover the LDOS, Eq.\,(\ref{eq:GUELDOS}), by integrating 
$R(E,\omega)$ over energy
\be
\frac{1}{N^2}\Omega(\omega)=\int \frac{dE}{4 \pi v^2} 
\Theta(2 E^2
+\frac{\omega^2}{2}-4v^2)
=
\frac{1}{\pi\sqrt{2v^2}}\sqrt{1-\frac{\omega^2}{8v^2}}
\nonumber
\,,
\ee
which serves also to verify that our analysis is internally
consistent.
\begin{figure}[h]
\epsfxsize=2.5truein
\centerline{\epsffile{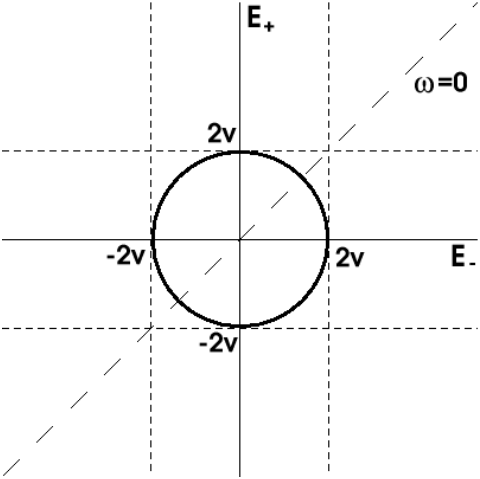}}
\caption{Analytic structure of $K(E_+,E_-)$. $R(E,\omega)$ is finite inside the circle.
}
\label{fig:thecross}
\end{figure}
\begin{figure}[h]
\epsfxsize=2.5truein
\centerline{\epsffile{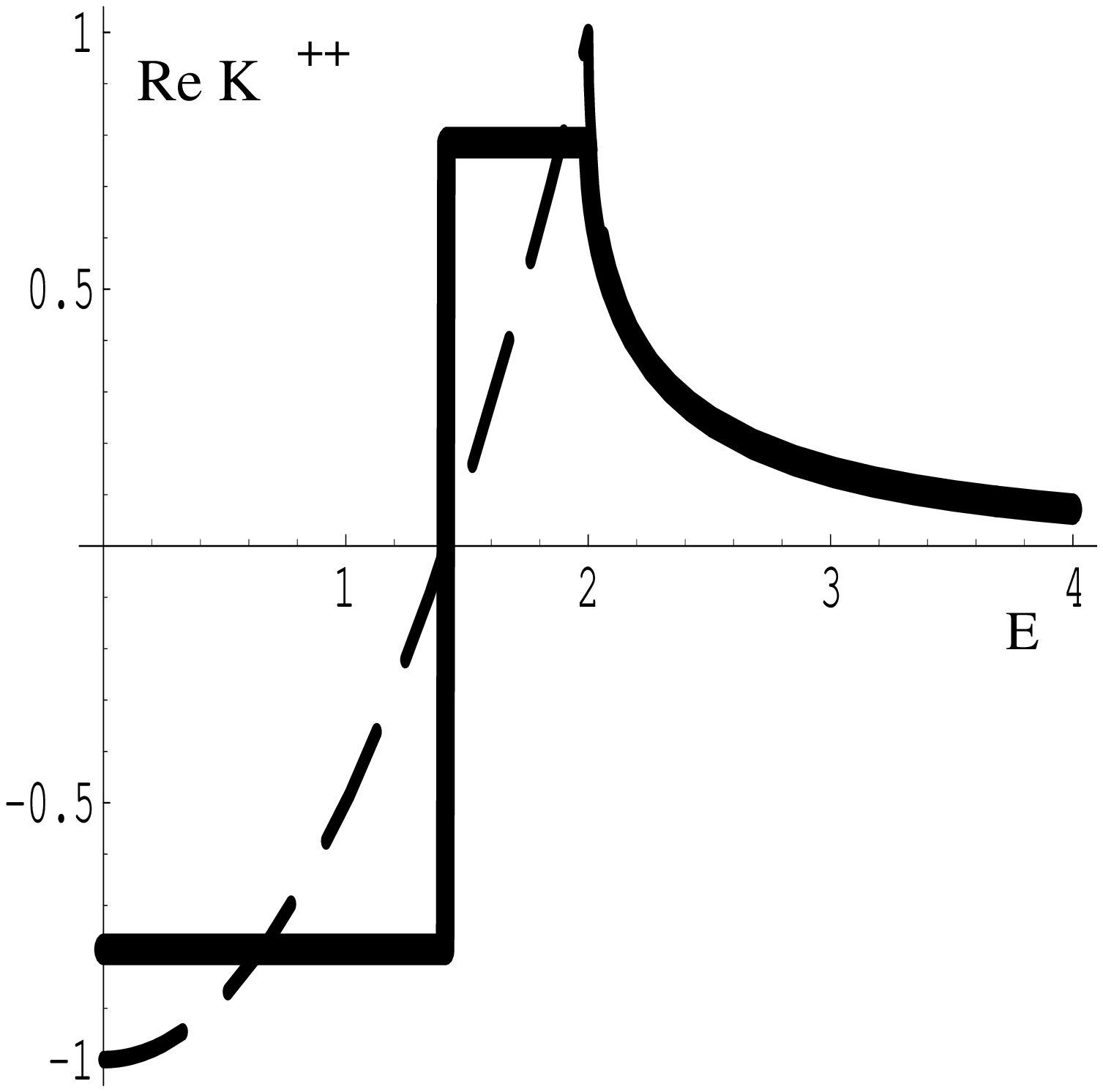}}
\caption{
Real part of $K^{++}(E,E)$ corresponding to the energy resolved Liouvillian solution (solid line) vs. exact result (dashed line).
}
\label{fig:ReK++}
\end{figure}
\begin{figure}[h]
\epsfxsize=2.5truein
\centerline{\epsffile{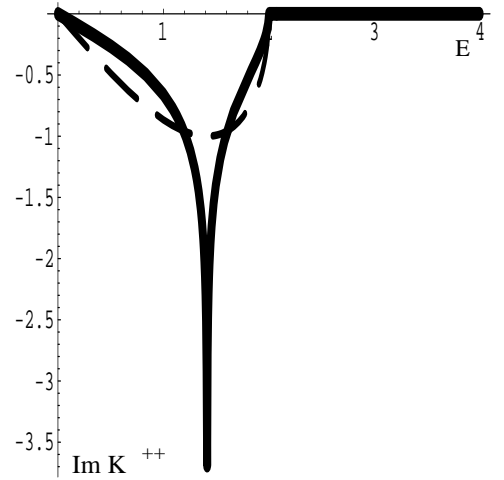}}
\caption{
Imaginary part of $K^{++}(E,E)$
corresponding to the energy resolved Liouvillian solution (solid line) vs. exact result (dashed line).
}
\label{fig:ImK++}
\end{figure}
\subsection{Summary of random matrix analysis}
Briefly, in our direct examination of the \lv method we find
that the large $N_L$ limit is unrepresentative of the
behavior of the $N_L=1$ problem. Upon energy unintegrating
the \lv $N_L=\infty$ calculation, we find a pathological approximation
which fails to factor correctly and even produces singularities
inside the band. Together, these facts indicate that the \lv
$1/N_L$ expansion is not a useful way of approaching 
the random matrix problem.

In this analysis we have utilized two complementary perspectives
 on the large $N_L$ method and a comment on those is perhaps
 useful. The first perspective is that we generate a family of
 problems with enlarged symmetry groups indexed by an integer
 each one of which can be studied directly---which is what we have 
 done in the numerical analysis. The enlarged problems are
 defined by multi-flavor Liouvillians that do not themselves
 arise from from single particle Hamiltonians.
 The second perspective treats 
 the large $N_L$ limit as formalizing perturbation theory about
 a saddle point that is believed to capture the relevant physics. 
 In this fashion we generate a series in powers
 of $1/N_L$ whose first term comes solely from the saddle point.
 The utility of the method is, of course, that setting $N_L=1$
 in the perturbation theory yields the answer for the case 
 of interest. In using the $N_L =\infty$ answer as input for
 the energy unintegration algorithm we have followed standard
 practice and simply truncated the series at its first term 
 and set $N_L=1$. If the large $N_L$ method is useful in
 describing the $N_L=1$ problem this should be a sensible
 approximation. Since we find that this is not so, the premise
 must be flawed.
\section{The Quantum Hall transition}
\label{sec:QH}
Our \lv analysis of the random matrix problem suggests to us 
that the method is unreliable and hence not to be
trusted in its application to the quantum Hall problem. This is
a disincentive to pursue a similarly detailed study
of the quantum Hall case, especially as it would be a substantial
undertaking: even the \lv $N_L = \infty$ analysis of that problem requires 
a numerical solution, and a full energy unintegration would require
much more intensive numerical work. We therefore confine ourselves to
three sets of observations. First, we confirm that our
general \lv formalism reduces to the one used by SMG and MSZ
for the quantum Hall case. Second, we exhibit the unintegration
algorithm specialized to this case. Third, we apply the
unintegration algorithm to a special sector of the $N_L = \infty$ problem,
that of very large momenta, where it reduces to precisely the
\lv random matrix problem studied in the last section. This then
imbeds the pathologies of that problem into the quantum Hall case.

\subsection{\lv formulation}

We consider a charged particle moving in a magnetic field
on a torus of area 
$2\pi l^2 N$, where $l \equiv ({\hbar c /e B})^{1/2}$ is the
magnetic length and $N$ is the integer degeneracy of each Landau level.
We denote the projection operator onto the lowest Landau level by ${\cal P}$. 
As the kinetic energy is constant within each Landau level, the 
Hamiltonian projected 
to the lowest Landau level is simply $H= {\cal P}V({\bf r}){\cal P}$ where
$V({\bf r})$ is the impurity potential.

It is convenient to work initially with the states
$\{{\cal P} |{\bf r}\rangle \}$ which form an overcomplete basis for
the lowest Landau level; to lighten notation we will write $\{|{\bf
r}\rangle \}$ 
for $\{{\cal P} |{\bf r}\rangle \}$. In terms of these we can define the
single-particle Green functions in real space 
\begin{equation}
G^{\pm}(E;{\bf r}_1,{\bf r}_2)=
\langle{\bf r}_1|\frac{1}{(E-H \pm i\delta)}|{\bf r}_2\rangle
\end{equation}
and the two-particle retarded-advanced Green function in real space 
\begin{equation}
K^{+-}(E, \omega;{\bf r_1},{\bf r_2}) =
G^{-}(E-\omega/2;{\bf r_1},{\bf r_2})
G^{+}(E+\omega/2;{\bf r_2},{\bf r_1})\,,
\end{equation}
where we have chosen a pairing of position coordinates
appropriate for a diffusion propagator. These are
evidently the analogs of Eqns.\,(\ref{eq:G}) and (\ref{eq:K})
in Section II. From these we derive the analog of Eq.\,(\ref{pidef}),
\ba
\Pi({\bf r_1}, {\bf r_2}, \omega) &=& 
\int_{-\infty}^{\infty} \, {dE \over 2\pi i}
K^{+-}(E,\omega;{\bf r_1},{\bf r_2}) \nonumber \\
&=&
({\bf r_1},{\bf r_1}| \frac{1}{( \omega +i\delta -{\cal L})}|{\bf r_2},{\bf r_2})\,.
\label{eq:Pi(r,0)}
\ea

As disorder averaging restores translational invariance, we take
Fourier transforms and define (denoting
the disorder averaged quantities by the same symbols),
\ba
&&K^{+-}(E,\omega;{\bf q}) = \nonumber \\
&&\frac{1}{2\pi l^2N}\int \int K^{+-}(E, \omega;{\bf r}_1,{\bf r}_2)
e^{i{\bf q}.({\bf r}_2-{\bf r}_1)}d{\bf r}_1d{\bf r}_2 \nonumber \\
\Pi(\omega;{\bf q}) &=& \frac{1}{2\pi l^2N}\int \int \Pi( \omega;{\bf r}_1,{\bf r}_2)
e^{i{\bf q}.({\bf r}_2-{\bf r}_1)}d{\bf r}_1d{\bf r}_2
\label{K2}
\ea
and the states
\begin{equation}
|{\rho}_{\bf q})=\int |{\bf r},{\bf r}) \exp(-i{\bf q}.{\bf r}) d^2{\bf r}\,,
\label{rho}
\end{equation}
which represent the projected density operators 
${\cal P}\exp(-i{\bf q}.{\bf r}){\cal P}$
in the space of operators $\cal A$ on the lowest Landau level. 
In terms of
these we can rewrite Eq.\,(\ref{eq:Pi(r,0)}) as
\ba
\Pi({\bf q},\omega)&=& \int_{-\infty}^{\infty} \, {dE \over 2\pi i}
K^{+-}(E,\omega; {\bf q})\nonumber \\
&=&
\frac{1}{2 \pi \hbar l^2 N} \langle({\rho}_{\bf q}|
\frac{1}{(\omega +i\delta -{\cal L})}|{\rho}_{\bf q})\rangle\,.
\ea

In place of the density operators, it is slightly more convenient to use 
(as a basis for $\cal A$) the magnetic translation operators 
$\tau_{\bf q} = e^{+l^2q^2/4} {\rho}_{\bf q}$,
which are closed under the algebra
\be
\tau_{\bf q} \tau_{\bf q'} = \exp({il^2 {\bf q} \wedge {\bf q}'/2}) \tau_{\bf q+q'}\,,
\ee
where $q \wedge q' = \epsilon_{ij} q_i q'_j$,
and are orthogonal with normalisation
\begin{equation}
(\tau_{{\bf q}1}|\tau_{{\bf q}2})=N\,\delta_{{\bf q}1,{\bf q}2}\,.
\end{equation}
With the identification
\begin{equation}
|{\rho}_{\bf q})=e^{-l^2q^2/4}|\tau_{\bf q})\,
\end{equation}
one can show that in this basis the matrix elements of $\hat{\cal L}=\hat{H} \circ
\hat{1}- \hat{1} \circ \hat{H}$ are
\ba
{\cal L}_{{\bf q}{\bf q}'} &=& V_{{\bf q}-{\bf q}'}
e^{-\frac{|{\bf q}-{\bf q}'|^2 l^2}{2}}
\left [e^{i\frac{{\bf q}\wedge {\bf q}'l^2}{2}}-
e^{-i\frac{{\bf q}\wedge {\bf q}'l^2}{2}}\right
]\\
&=&
2i V_{{\bf q}-{\bf q}'} e^{-\frac{|{\bf q}-{\bf q}'|^2 l^2}{2}}
\sin(\frac{{\bf q} \wedge {\bf q}' l^2}{2})\,  .
\ea
With these we arrive finally at the form
\begin{equation}
\Pi({\bf q},\omega)= \frac{\exp(-q^2l^2/2)}{2 \pi l^2}
\left<
\left[\frac{1}{\omega +i\delta -\hat{\cal L}}\right]_{{\bf q}{\bf q}}\right>\,,
\end{equation}
which is the one introduced by SMG (up to a constant due to difference
in definitons).  

From known facts about the two particle Green function it can be
deduced (see Appendix A for details) that the exact \lv self energy
varies for small $q,\omega$ as
\ba
\omega - \Pi^{-1}({\bf q},\omega)\equiv i q^2 D(\omega)
\sim i q^2 \omega^{\frac{1}{2\nu}}.
\ea
This is sub-diffusive behavior and is controlled by the 
correlation length exponent 
as $q \rightarrow 0$ at fixed $\omega$. The combined work of SMG and
MSZ has yielded a self energy, in the $1/N_L$ expansion, of this form 
with
\be
D(\omega)\sim D_0[1 + O( {\log \omega \over N_L} )]\,,
\ee
suggestive of an expansion of a critical power law. In the following
subsection we note that the general algorithm of Section II
could be used to examine this result at different energies in the
disorder broadened Landau level.

\subsection{Energy unintegration}

Formally, the procedure for recovering $K({\bf q},E,\omega)$ from
$\Pi({\bf q},\omega)$ is a straightforward specialization of
Section  \ref{sec:gendef}. One starts by deforming $\hat{\cal L}$ to
\ba
{\cal L}_{{\bf q}{\bf q}'}(p,h) &=& V_{{\bf q}-{\bf q}'}
e^{-\frac{|{\bf q}-{\bf q}'|^2 l^2}{2}}
\left [p e^{i\frac{{\bf q}\wedge {\bf q}'}{2}l^2}
-h e^{-i\frac{{\bf q}\wedge {\bf q}'}{2}l^2}\right
] \ . \nonumber
\ea
This generates a deformation of any given \lv diagram or of some
partial summation 
that leads to an approximate form for $\Pi({\bf q},\omega)$. 
In either case, carrying out the
replacements in Eq.\,(\ref{eq:Piph}) and the integrals
in Eq.\,(\ref{eq:triple}) gives the corresponding
energy resolved expressions. 

As for the general example of Sec. \ref{sec:gendef}
 the algorithm relies on disorder 
contractions to keep track of the topology of individual diagrams with
$p$ 
and $h$ used to assign appropriate weights via Eq. \ref{eq:triple}. Since the
first 
step of the procedure is particularly transparent in the LL setting we 
briefly consider the analog of Fig. \ref{fig:HHHH}.
 In particular, for each of 
the ten ``unfolded'' diagrams we can immdediately write down its value
by drawing momentum carrying lines (both disorder and external) inside 
the ``particle-hole circle'' and integrating correlators of disorder
potential with an exponential/phase factor chosen according to diagram's
topology. For example, 
the diagram of Fig. \ref{fig:wedges} is given (aside from energy
dependence) 
by 
\begin{figure}
\epsfxsize=2.5truein
\centerline{\epsffile{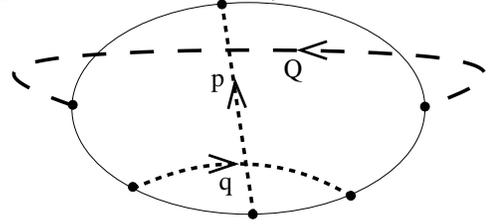}}
\caption{
One of the ten diagrams Fig. \ref{fig:HHHH} redrawn and relabelled using
momentum coordinates.
}
\label{fig:wedges}
\end{figure}
\be
e^{-\frac{Q^2 l^2}{2}}\int \frac{d^2 q d^2 p}{(2 \pi)^4}\langle V_{\bf
q}V_{\bf -q}\rangle 
\langle V_{\bf p}V_{\bf -p}\rangle e^{-\frac{(p^2+q^2)
l^2}{2}-i({\bf q}\wedge {\bf p} +{\bf p}\wedge{\bf Q})l^2},
\ee
and boils down to evaluating a determinant.
The wedge factors (${\bf q}\wedge {\bf p}\equiv {\bf\hat{z}}\cdot
{\bf q}\times {\bf p}$) appear whenever two momentum lines cross (
one also selects a sign convention) and effectively serve as a
fingerprint of each diagram. These diagrammatic rules
are familiar to workers in the field and have enabled considerable
progress in series analysis of the problem \cite{singh,wegner}.
Here, they illustrate the connection between \lv and Green function
perturbation series.

In order to apply this procedure in full to the \lv $N_L = \infty$ limit,
we must compute $\Pi({\bf q},\omega,p,h)$, which 
in turn requires numerical solution of an integral equation and is therefore
a substantial undertaking. We turn now to the large ${\bf q}$
limit, for which the energy unintegration can be carried out more easily.

\subsection{Energy unintegration at 
$N_L = \infty$, as ${\bf q} \rightarrow \infty$}

The key observation\cite{MZS} that allows us to execute the 
energy unintegration procedure
in this limit is that the self-consistency equation simplifies drastically.
Defining $\overline{\Pi}({\bf q},\omega)$ by 
$\Pi({\bf q},\omega) = e^{-l^2 q^2/2}\overline{\Pi}({\bf q},\omega)$,
as  $q\rightarrow \infty$ the self-consistency equation
\begin{widetext}
\ba
\overline{\Pi}({\bf q},\omega)&=&\frac{1}{\omega+i \delta}+
\frac{\overline{\Pi}({\bf q},\omega)}{\omega + i\delta} \int \frac{d^2
q'}{4\pi^2}
\langle\hat{\cal L}_{\bf q,\bf q+\bf q'}
\hat{\cal L}_{\bf q+ \bf q',\bf q} \rangle
\overline{\Pi}({\bf q} + {\bf q'},\omega)
\ea
\end{widetext}
reduces to  
\ba
\overline{\Pi}(\infty,\omega)&=&\frac{1}{\omega+i\delta}+\frac{2v^2
\overline{\Pi}^2 (\infty,\omega)}{\omega+i \delta},
\ea
which is {\it identical} to the corresponding RMT equation 
(either GUE or GOE)
at $N=\infty$. It follows then that the energy unintegrated solution of
this equation, the 
quantity  $\overline{K}({\bf \infty},E_+,E_-) = \lim_{\bf q \rightarrow
\infty} e^{+q^2 l^2/2} K({\bf q},E_+,E_-)$,  is given by the expression
already exhibited in Eq.\,(\ref{eq:Kunint}):
\begin{widetext}
\ba
{\overline K}({\bf \infty},E_+,E_-)&=&
\frac{1}{2v^2}\left[\sin^{-1}\frac{E_+ \sqrt{E_+^2-4v^2}+E_-
\sqrt{E_-^2-4v^2}}{E_{+}^2+E_{-}^2-4v^2}
-\frac{\pi}{2}\right].
\ea
\end{widetext}
As before, the two particle Green function fails to factorize in
energy. This has, however, a more serious consequence in the
quantum Hall problem. Large momenta also correspond to large
spatial separations in magnetic fields. More precisely, one has
\begin{widetext}
\ba
\langle G^+({\bf r},{\bf r},E_+)G^-({\bf 0},{\bf 0},E_-) \rangle
&=&\int 
\frac{d^2 q}{4\pi^2} \exp({-\frac{1}{2l^2}|{\bf z}\times{\bf r} + {\bf q}
l |^2}) \overline{K}({\bf q},E_+,E_-),
\ea
and hence
\be
\lim_{{\bf r} \to \infty}\langle G({\bf r},{\bf r},E_+)G({\bf 0},{\bf 0},E_-) \rangle
=\lim_{{\bf q} \to \infty}\frac{1}{2\pi l^2} \overline{K}({\bf q},E_+,E_-)\,.
\ee
\end{widetext}
So we discover that lack of energy factorization 
translates into a lack of factorization at infinite spatial separation, which
is clearly unphysical. In addition, we again find that the 
retarded-retarded Green function is
singular inside the impurity band in a limit where the exact answer is
neccessarily analytic.

\section{Summary and Outlook}
As this has been a largely technical discussion of the Liouvillian
formalism, it is perhaps useful to summarize the main argument again.
The work of SMG and MSZ has suggested that the critical divergence
of the localization length for the (non-interacting) quantum Hall
transition can be computed within the Liouvillian formalism by
the $1/N_L$ expansion in the number of \lv flavors.
We find, based on testing the \lv approach on random martrix theory
and energy unintegrating it at large momenta in the quantum Hall
problem, that this program has serious problems already at $N_L =
\infty$ which do not appear to be remediable within the
$1/N_L$ expansion. Accordingly, we conclude that the calculation
of MSZ does not represent a computation of the quantum Hall
correlation length exponent. As there isn't another approximation
scheme that readily suggests itself in the \lv formalism,
we are pessimistic, as indicated by our title, about its
utility for making progress on the problem of the quantum Hall
transition.

That said, there are a couple of loose ends in our analysis
that would be nice to tie up.
First, we have not carried out the energy unintegration of the
\lv  $N_L = \infty$ answer for the quantum Hall problem near
${\bf q} = 0$ and it would be of some technical interest to
see if that has an analytic structure different from large
momenta as well as a technical challenge to see how that might
get done. Second, it may be possible to use our ideas on
energy unintegration to see if the logarithm found by
MSZ has any interpretation in Green function perturbation
theory. Perhaps an intrepid reader will be inspired to
sort these out.

Finally, as we were finishing this work, there appeared a preprint\cite{JEM}
by Moore which studies the evolution of the QH \lv as a function of $N_L$.
Its most relevant finding, via numerical analysis at small $N_L$,
is that the \lv theory likely exhibits metallic diffusion, instead of critical
scaling, at all $N_L>1$. Evidently, this is consistent with our own
conclusion in Section \ref{sec:RMTnumerics} that the theory for all
$N_L{\not =}1$ is unrepresentative of the original problem of
interest ($N_L=1$).

{\it Acknowledgements:} We have benefitted from numerous invigorating conversations with E. Fradkin, S. Girvin, S. Sachdev, I. Ussishkin 
and, especially, J. Moore. This research was supported in part by the National Science Foundation through grant No. DMR 99-78074 (VO and SLS), by the David and Lucille Packard Foundation (VO and SLS), and by the EPSRC under grant GR/J78327 (JTC).
\appendix
\section{Critical behavior of the quantum Hall Liouvillian}
For completeness, but also because some of the original
discussion\cite{SMG} was incorrect\cite{Dq}, we discuss here how it
is that the energy integrated \lv propagator can be used to
extract the localization length exponent.
One might worry that, as the critical wavefunctions are a set of 
measure zero in the spectrum, any information on them will be lost upon 
integrating over the band and the behavior at small $\omega$ and $q$ 
will be that of an insulator. This is not the case: the \lv Green 
function exhibits critical asymptotics intermediate between
those of the metal and insulator in the transport limit $q \ll \omega$,
with the diffusion constant $D(\omega, {\bf q=0})
\sim \omega^{{1}/{2\nu}}$,  where $\nu$ is
the correlation length exponent\cite{Dq}.

Near the critical energy $E_c$, the two particle Green function for the
quantum Hall problem can be parametrized as \cite{chalker-param} 
\be
K(E, \omega ; {\bf q})=
\frac{2 \pi \rho(E)}{i \omega - D(E,\omega ; {\bf q}) q^2} \ .
\label{k-param}
\ee
In the transport limit $D q^2 \ll \omega$ we can expand the denominator
to obtain
\be
K(E, \omega ; {\bf q}) \approx 2 \pi \rho(E) \left[ {1 \over i \omega}
- {D q^2 \over \omega^2} \right] \ .
\label{kexp}
\ee
These forms also hold for a metal, with $D(E, \omega ;{\bf q = 0})$ non-zero,
and for an insulator with $D(E, \omega ; {\bf q = 0})$ vanishing 
for $\omega \to 0$ roughly
as $\omega^2$. For the quantum Hall problem, $D(E,\omega,{\bf q})$ 
has a non-trivial dependence
on $E$, $\omega$ and $\bf q$.

To quantify this dependence, consider the two important length scales
in the transport limit. These are the 
localization length $\xi \sim l | {E_c /(E - E_c)} |^{\nu}$ and
$L_\omega=(\rho(\epsilon) \omega)^{-1/2}$ 
which is the  size of a box with level 
spacing of order the frequency.
Then, when $L_\omega \ll \xi (E) $, localization is unimportant and we may take
$D(E,\omega,{\bf q}) \approx D_c$, a non-zero constant. In the opposite limit, 
$L_\omega \gg \xi (E)$,
we have the insulating behavior, $D(E,\omega,{\bf q})=D_{\rm ins}(\omega) 
\sim \omega^2$.
In the following we will assume a sharp crossover between these two limiting forms,
which is sufficient for our purposes.

Armed with these facts we can now carry out the energy integration to 
deduce the form of $\Pi({\bf q}, \omega)$ in the transport limit.
Specifically,
\be
\int dE {\rm Re} K(E, \omega ; {\bf q}) =
2 \pi \frac{q^2}{\omega^2} \int_{-\infty}^{\infty} 
dE \rho(E) D(E,\omega,{\bf q = 0})  \ .
\ee
We will see that the integral is dominated by energies near $E_c$ so
that we can replace $\rho(E) \approx \rho(E_c)$ and trade the 
integral over $E$ for one over the localization length using the
asymptotic relation $\xi \sim l | {E_c /( E - E_c}) |^{\nu}$.
This yields
\begin{widetext}
\ba
{\rm Im} \Pi({\bf q}, \omega)& \sim &
\frac{q^2}{\omega^2} {E_c l^{1/\nu} \over \nu} \rho(E_c)
\int_{l}^{\infty} d\xi \xi^{-1-1/\nu} 
D( E_c + E_c ({l / \xi})^{1/\nu},\omega,{\bf q = 0}) \nonumber \\
&\sim& \frac{q^2}{\omega^2} {E_c l^{1/\nu} \over \nu} \rho(E_c)
\left[ \int_{l}^{L_\omega} d\xi \xi^{-1-1/\nu} D_{{\rm ins}}(\omega)
+ \int_{L_\omega}^{\infty} d\xi \xi^{-1-1/\nu} D_{c}\right] \nonumber\\
&\sim&  \frac{q^2}{\omega^2} E_c  \rho(E_c)
\left[D_{{\rm ins}}(\omega)+D_c (l^2 \rho(E_c)\omega)^{{1}/{2\nu}}\right] 
\sim \frac{D_c q^2}{\omega^2}  (l^2 \rho(E_c)\omega)^{{1}/{2\nu}} \ ,
\ea
\end{widetext}
where the last simplifcation follows upon noting 
that $D_{{\rm ins}}(\omega)\sim \omega^2$
is much smaller than $ \omega^{\frac{1}{2\nu}}$ for $\nu>1/4$, which is
true in general of a random critical point in two 
dimensions on the grounds of the Harris criterion,
and holds for the reasonably precise estimates of $\nu$ available in this case.
As a statement about Liouvillian theory, this result 
implies that the self-energy of the exact Liouviliian Green function 
has the behavior 
$\Sigma(\omega,q \rightarrow 0)=\omega-1/\Pi(\omega,q)\sim i q^2 
\omega^{\frac{1}{2\nu}}$. 

\begin{widetext}
\section{Details of Energy Unintegration}
We start from Eq.\,(\ref{eq:unint}). After a change of variables, it reads
\ba
K(E_+,E_-)&=&
\int_0^\frac{\pi}{2} d\theta \int_0^\infty \frac{rdr \exp({-r (E_-
\cos\theta+E_+ \sin\theta))}}{2v^2 r^2} \oint \frac{dz e^z}{2\pi
i}[z-\sqrt{z^2-(2vr)^2}] \nonumber \\
&=&\int_0^\frac{\pi}{2} d\theta \int_0^\infty \frac{dr \exp({-r (E_-
\cos\theta+E_+ \sin\theta))}}{v} I_1(2 v r) \nonumber \\
&=&
\frac{1}{2v^2}\int_0^{\frac{\pi}{2}}d\theta\left(
\frac{R\cos(\theta+\phi)}{\sqrt{R^2\cos^2(\theta+\phi)-1}}-1\right), 
\ea
where we have defined 
\ba
R&\equiv&\frac{1}{2v}\sqrt{E_+^2+E_-^2}
=\frac{1}{2v}\sqrt{E^2+({\omega}/{2}+i 0^+)^2}
\ea
and
\ba
\tan \phi&\equiv& \frac{E_-}{E_+} \ .
\ea
Proceeding further,
\ba
K(E_+,E_-)
&=&
\frac{1}{2v^2}\left[\sin^{-1}\frac{R\cos\phi}{\sqrt{R^2-1}}+\sin^{-1}\frac{R\sin\phi}{\sqrt{R^2-1}} -\frac{\pi}{2}\right]\nonumber \\ 
&=&
\frac{1}{2v^2}\left[\sin^{-1}\frac{E_+}{\sqrt{E_{+}^2+E_{-}^2-4v^2}}+\sin^{-1}\frac{E_-}{\sqrt{E_{+}^2+E_{-}^2-4v^2}}
-\frac{\pi}{2}\right] \nonumber\\
&=&
\frac{1}{2v^2}\left[\sin^{-1}\frac{E_+ \sqrt{E_+^2-4v^2}+E_-
\sqrt{E_-^2-4v^2}}{E_{+}^2+E_{-}^2-4v^2}
-\frac{\pi}{2}\right]
\ea
which is Eq.\,(\ref{eq:Kunint}).
\end{widetext}
\section{Magnetic algebra and random matrices}
In this appendix we demonstrate that an algebra like that of density
operators in the lowest Landau level may be set up for an arbitrary
Hamiltonian.
We consider an orthonormal basis $\{|a\rangle\}$ for ${\cal V}$
and the corresponding basis $\{|a,b\rangle\}$ on ${\cal A_{\cal V}}$, 
where $a$ and
$b$ range from $1$ to $N$).
We define 
\be
|{\bf q})=\frac{1}{N^{1/2}}e^{i q_x q_y \pi/N}\sum_n
|n+q_y,n)
 e^{i q_x n 2\pi/N}
\ee
 for integer $q_x$ and $q_y$ (ranging from $1 \to N$) which also form
an orthonormal basis.
We can rewrite
\be
\hat{H}=\sum_{mn}H_{mn}|m,n)=\sum_{\bf q}V_{-{\bf q}}|{\bf q}),
\ee
where
\be
V_{-{\bf q}}\equiv \sum_{n} \frac{H_{n+q_y,n}}{N^{1/2}}e^{-i q_x (n+q_y/2)2\pi/N}
\ee
Switching to operator notation, with $\tau_{\bf q} \equiv |{\bf q})$,
one can check that 
\be
[\tau_{\bf q},\tau_{\bf p}]=2i\sin \frac{\pi {\bf q}\wedge {\bf p}}{N}
\tau_{{\bf q}+{\bf p}}
\ee
so that 
\be
\frac{\partial}{i\partial t}|{\bf q})
=\frac{1}{N^{1/2}}\sum_{{\bf p}}2 i V_{-{\bf p}} \sin \frac{\pi {\bf
q}\wedge {\bf p}}{N}|{\bf p}+{\bf q}).
\ee
These expressions parallel those for the \lv in the lowest
Landau level. Strikingly we have recast a generic 
problem in the notation of the magnetic translation
operators.
Of course, the definition of the (discrete) momenta is
adhoc, and so, while this rewriting appears suggestive, it is not
generally useful. Nevertheless, the GUE retains its simplicity in this
approach: Eq.\,(\ref{eq:HH}) 
implies that 
\be
\langle V_{\bf p} V_{\bf q}\rangle=\frac{v^2}{N} \delta_{{\bf p},-{\bf q}}.
\ee
Disorder-correlators are thus diagonal in ${\bf q}$, 
continuing the analogy with spatially extended systems.


\begin{references}
\bibitem{Qinfinity}
One easy way to see that this is true is by assuming no dependence at large momenta and verify self-consistency of this assumption. More systematicly, the corrections 
are Gaussianly small.
\bibitem{exact2}
LDOS can be computed exactly for $N=2$ and $N_L=1$:
\ba
{\rm GOE:}&&\Omega(\omega)=2
\delta(\omega)+\frac{|\omega|}{v^2} e^{-\frac{\omega^2}{2 v^2}} \\
{\rm GUE:}&&\Omega(\omega)=2
\delta(\omega)+\frac{4\omega^2}{\sqrt{\pi} v^3}  e^{-\frac{\omega^2}{v^2}}
\ea
Note that Eq. (\ref{eq:HH}) is equivalent, for general $N$, 
to  matrices being distributed according to 
${\cal P}[\hat{H}] \sim exp[-\frac{N}{(1+\alpha)v^2} Tr\{\hat H^2\}]$.
\bibitem{Dq}
SMG proposed instead a wavevector dependent $D(q)\sim q^{\frac{1}{\nu}}$. This was corrected later by Gurarie and Zee\cite{GurZee}.
\end{references}
\end{document}